\documentclass[fleqn,usenatbib,onecolumn]{mnras}

\usepackage{newtxtext,newtxmath}
\usepackage{graphicx}

\usepackage{amsmath,bm}
\usepackage{lscape}

\usepackage[T1]{fontenc}

\DeclareRobustCommand{\VAN}[3]{#2}
\let\VANthebibliography\thebibliography
\def\thebibliography{\DeclareRobustCommand{\VAN}[3]{##3}\VANthebibliography}

\def \aj {{\rm {AJ}}}

\def \apj {{\rm {ApJ}}}

\def \apjs {{\rm {ApJS}}}
\def \procspie {{\rm {Proc. SPIE}}}

\def \aap {{\rm {A\&A}}}

\def \mnras {{\rm {MNRAS}}}

\def \apjl{\rm {ApJL}}
\def \nat{\rm {Nat.}}

\def \pasp{\rm {PASP}}

\def \mj{\,$M_{\rm Jup}$\,}

\title[Lessons learned from the detection of  companions]{Lessons learned from the detection of wide companions by radial velocity and astrometry}
\author[F. Feng et al.]
{Fabo Feng$^{1}$\thanks{E-mail: ffeng@sjtu.edu.cn}, Guang-Yao Xiao$^{1}$\thanks{E-mail: gyxiao\_tdli@sjtu.edu.cn}, Hugh R. A. Jones$^{2}$, James S. Jenkins$^{3, 4}$, Pablo Pena$^{3,4}$,
\newauthor
Qinghui Sun$^{1}$\\
$^{1}$State Key Laboratory of Dark Matter Physics, Tsung-Dao Lee Institute \& School of Physics and Astronomy, Shanghai Jiao Tong University, Shanghai 201210, China\\
$^{2}$Centre for Astrophysics Research, University of Hertfordshire, College Lane, AL10 9AB, Hatfield, UK\\
$^{3}$Instituto de Estudios Astrof\'isicos, Facultad de Ingenier\'ia y Ciencias, Universidad Diego Portales~Santiago~ Av. Ej\'ercito 441\\
$^{4}$Centro de Astrof\'isica y Tecnolog\'ias Afines (CATA) ~Santiago~Casilla 36-D\\
}
\date{\today}

\begin{document}
\maketitle
\begin{abstract}
The detection and constraint of the orbits of long-period giant planets is
essential for enabling their further study through direct
imaging. Recently, it has been highlighted that there are discrepancies between different orbital fitting solutions.
 We address these concerns by reanalyzing the data for HD 28185, GJ 229, HD 62364, HD 38529, 14 Her, $\epsilon$ Ind A, HD 211847, HD 111031, and GJ 680,
offering explanations for these discrepancies. Based on the comparison between our direct modeling of the astrometric catalog data and the \texttt{orvara} code,  we find the  discrepancies are primarily data-related rather than methodology-related. Our re-analysis of HD 28185 highlights many of the data-related issues and particularly the importance of parallax modeling for year-long companions. The case of eps Ind A b is instructive to emphasize the value of an extended RV baseline for accurately determining orbits of long period companions. Our orbital solutions highlight other causes for discrepancies between solutions including the combination of absolute and relative astrometry, clear definitions of conventions, and efficient posterior sampling for the detection of wide-orbit giant planets.
\end{abstract}
\section{Introduction}\label{sec:introduction}
The detection and characterization of cold Jupiters have become feasible with high-precision data from radial velocity (RV) facilities, Gaia astrometry, and the imaging by the mid-infrared instrument (MIRI; \citealt{rieke15}) installed on the James Webb Space Telescope (JWST). As RV, astrometry, and imaging techniques all target planets around nearby stars, they can be used synergistically to gather more comprehensive information than each technique would individually provide.

Since Gaia's second data release, several research groups have developed methods for jointly analyzing Hipparcos and Gaia data \citep{snellen18,feng19b, kervella19,li21,xiao23}. Among the many orbital fitting packages, such as \texttt{Octofitter} \citep{thompson23}, \texttt{GaiaPMEX} \citep{kiefer24}, and \texttt{Nii-C} \citep{jin24}, one widely used tool is \texttt{orvara} \citep{brandt21c}. This package constrains reflex motion using the Hipparcos-Gaia Catalog of Accelerations (HGCA; \citealt{brandt18,brandt21}). HGCA provides three proper motion values from the Hipparcos and Gaia catalogs, including the proper motions at the Hipparcos and Gaia reference epochs and the mean proper motion. Combined with \texttt{htof} \citep{brandt21a}, \texttt{orvara} can model the reflex motion’s orbit using Gaia epoch data from Gaia GOST and Hipparcos Intermediate Astrometric Data (IAD).

While \cite{brandt19} and \cite{brandt21} transform the calibrated Hipparcos and Gaia catalog data into proper motions, \cite{feng19b} (hereafter F19) and \cite{feng22} (hereafter F22) directly model the astrometric catalog data. They incorporate five parameters for barycentric motion and seven parameters for reflex motion, astrometric jitter, and offsets to account for potential biases in the catalog data. The F19 method was later enhanced to model both Gaia DR2 and DR3 data \citep{feng23} (hereafter F23), utilizing the Gaia Observation Forecast Tool (GOST). RV data primarily constrain five orbital parameters, while astrometric data help determine the inclination ($I$) and longitude of the ascending node ($\Omega$). Instead of calibrating the Hipparcos and Gaia data a priori, the F19 and F23 methods use astrometric jitters and offsets to account for potential biases in the Hipparcos catalog and infer these parameters a posteriori.

Recently, \cite{venner24} (hereafter V24) argued that the method developed by F19 and applied in wide companion detections by F22 is unreliable, citing discrepancies between F22 and other studies. This paper addresses V24's criticisms, contending that these discrepancies stem not from methodological issues but from 1) differences in datasets and conventions, 2) insufficient posterior sampling, 3) partial radial velocity coverage, and 4) the influence of short-period companions. With partial RV coverage of the orbital phase of long-period planets, degeneracies may arise between mass, orbital period, and eccentricity. These degeneracies can be resolved by incorporating relative astrometry, extending the RV baseline, and ensuring adequate posterior sampling.

This paper is structured as follows: In Section
\ref{sec:comparison}, we compare the F19 method and its updated implementation with \texttt{orvara}. Next, we reanalyze the HD28185
data and compare our findings with those in V24 (Section
\ref{sec:HD28185}). We then address discrepancies between F22 and other studies in Section \ref{sec:discrepancy}. We also examine the case of eps Ind A b to highlight key lessons in Section \ref{sec:epsIndA}. Finally, we present our discussion and conclusions in Section \ref{sec:conclusion}.

\section{Comparison of different methods}\label{sec:comparison}
Upon examining the differences between \texttt{orvara} and the F19 method, we agree with V24 that the two approaches are nearly equivalent. The key distinction is that F19 uses astrometric jitter to account for systematic bias, while \texttt{orvara} employs calibrated HGCA data. We define ${\bm r}_H$ and ${\bm r}_G$ as the reference position vectors for Hipparcos and Gaia, respectively, with ${\bm \mu}_H$ and ${\bm \mu}_G$ as the corresponding proper motions, and ${\bm \mu}_{HG}$ as the mean proper motion, calculated as $({\bm r}_G - {\bm r}_H) / \Delta T$, where $\Delta T = t_G - t_H$ is the interval between Hipparcos and Gaia reference epochs. Both positions and proper motions are utilized in the F19 method.

For convenience, we define ${\bm r}_0$ and ${\bm \mu}_b$ as the
barycentric position at the Hipparcos reference epoch and the
barycentric proper motion. Additionally, we define ${\bm \mu}^r_X$ and
${\bm r}^r_X$ as the proper motion and position at epoch $X$, which
could be either Gaia (G) or Hipparcos (H) epochs. The position model
is thus:
\begin{align}
  {\bm r}_H&={\bm r}_0+{\bm r}^r_H~,\\
  {\bm r}_G&={\bm r}_0+{\bm\mu}_b\Delta T+{\bm r}^r_G~.
\end{align}
The difference ${\bm r}_{G} - {\bm r}_{H}$, when divided by $\Delta T$,
yields:
\begin{equation}
  \hat{\bm\mu}_{HG}\equiv\frac{{\bm r}_{\rm G}- {\bm r}_{\rm H}}{\Delta
    T}={\bm \mu}_{\rm b}+\frac{{\bm r}^r_{\rm G}-{\bm r}^r_{\rm
      H}}{\Delta T}~.
\end{equation}
By setting ${\bm \mu}_{HG} \equiv \frac{{\bm r}^r_{G} - {\bm
    r}^r_{H}}{\Delta T}$, we obtain:
\begin{equation}
  \hat{\bm\mu}_{HG}={\bm \mu}_b+{\bm \mu}_{HG}~,
\end{equation}
matching the model for observed mean proper motion $\mu_{HG,o}$ as defined by \cite{brandt21c}. Both F19 and \texttt{orvara} model Gaia and Hipparcos proper motions, ${\bm \mu}_G$ and ${\bm \mu}_H$, similarly, making them equivalent for orbit constraints, though F19 can also yield barycentric positions at reference epochs. Thus, the difference between F19 and \texttt{orvara} lies solely in the parameterization and not in the physical content.

\begin{figure*}
  \begin{center}
    \includegraphics[scale=0.55]{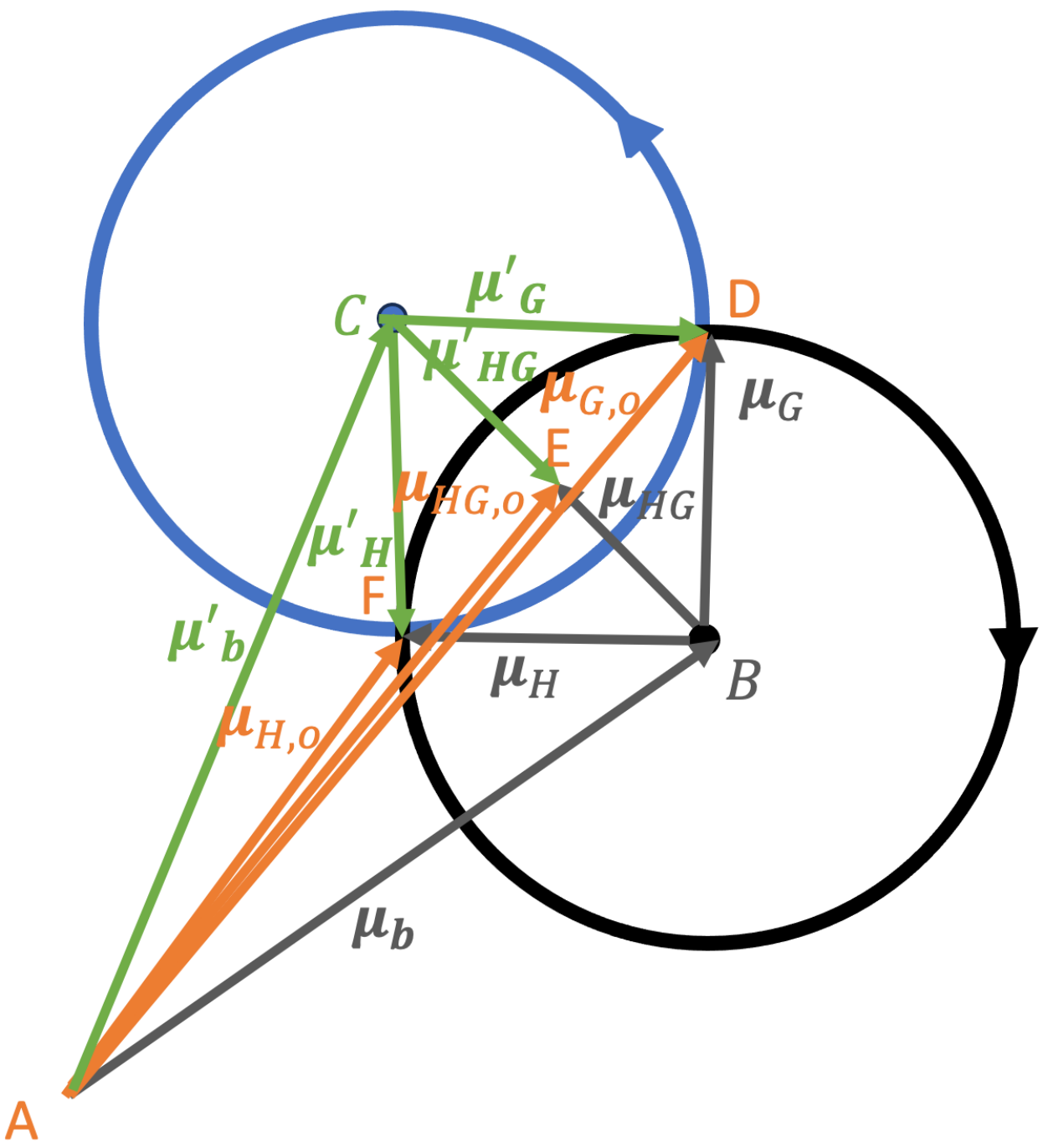}
    \caption{Schematic view of the two orbital solutions in proper motion space. The black and blue circles with directed arrows represent two equivalent orbital solutions. Grey and green vectors denote the proper motion vectors associated with stellar reflex motion, constrained by the observed proper motion vectors shown in orange. Point $A$ marks the origin of the coordinate system, while $B$ and $C$ represent the barycenters for the two solutions. The vectors $\protect\overrightarrow{AD} \equiv {\bm \mu}_{G,o} $ and $ \protect\overrightarrow{AF} \equiv {\bm r}_{H,o} $ denote the observed proper motions at the Gaia and Hipparcos reference epochs, respectively, and $ \protect\overrightarrow{AE} \equiv {\bm \mu}_{HG,o} $ represents the observed mean motion calculated from the positional difference between Hipparcos and Gaia. $ \protect\overrightarrow{BD} \equiv {\bm \mu}_G $ and $ \protect\overrightarrow{CD} \equiv {\bm \mu}_H $ indicate the stellar position relative to the barycenter at the Gaia and Hipparcos reference epochs, respectively. $ {\bm \mu}_{HG} $ is the proper motion derived from the positions of the reflex motion at the Gaia and Hipparcos epochs. Quantities with a prime symbol correspond to the alternate solution.}
\label{fig:two_solution}
\end{center}
\end{figure*}

Both methods share a limitation in that they cannot resolve the degeneracy between $\pi - I$ and $I$, leading to two possible inclination solutions, one for a retrograde and one for a prograde orbit. This ambiguity is illustrated in Fig. \ref{fig:two_solution} by two face-on circular orbits. In \texttt{orvara}, the observed quantities ${\bm \mu}_{H,o}$, ${\bm \mu}_{G,o}$, and ${\bm \mu}_{HG,o}$ help constrain the reflex motion components ${\bm \mu}_{H}$, ${\bm \mu}_{G}$, and ${\bm \mu}_{HG}$ as well as the barycentric proper motion ${\bm \mu}_b$. For a circular orbit, where the orbital period is primarily constrained by RV data, the radius of the orbital path is fixed. Thus, for a circular and face-on orbit, ${\bm \mu}_{H,o}$ and ${\bm \mu}_{G,o}$ primarily determine the direction of orbital motion, while ${\bm \mu}_{HG,o}$ constrains ${\bm \mu}_{HG}$. With two identical orbital circles that intersect two fixed points (i.e. D and F in Fig. \ref{fig:two_solution}), two orbital solutions can result, representing either rotation direction when using \texttt{orvara} or F19 methods.

To resolve the inclination or rotation degeneracy, one can utilize data from both Gaia DR2 and DR3, along with Hipparcos, to better constrain orbits. In the simplest case illustrated in Fig. \ref{fig:two_solution}, a circular orbit can be uniquely determined with three fixed
points in 2D space. As shown in Fig. \ref{fig:two_solution}, an additional point is needed to distinguish between the two solutions represented by the blue and black circles. This approach is demonstrated by comparing the orbital
solutions for HD 222237 b provided by the F23 and \texttt{orvara}
methods \citep{xiao24}. An additional advantage of the F23
method over both F19 and \texttt{orvara} is its ability to model the
parallax contribution to abscissae using GOST, while simultaneously fitting both barycentric motion and reflex motion
models to data from Hipparcos and multiple Gaia data releases.

\section{Orbital Analysis of HD 28185}\label{sec:HD28185}
HD 28185 is a G-type star \citep{gray06} hosting at least two substellar companions, designated HD 28185 b and c. While prior studies, including V24 and F22, concur on the minimum masses and orbital parameters for the inner planet HD 28185 b \citep{santos01,minniti09,wittenmyer09,stassun17,rosenthal21}, they differ in their interpretations of HD 28185 c. F22, using a combined analysis of RV and Hipparcos-Gaia astrometry, arrives at an orbital solution for HD 28185 c similar to that based on RV data alone from \cite{rosenthal21}. V24, however, presents a contrasting solution, with significant differences highlighted in fig. 4 of V24. This discrepancy likely arises from two main factors.

\begin{figure*}
    \centering
	\includegraphics[width=0.88\textwidth]{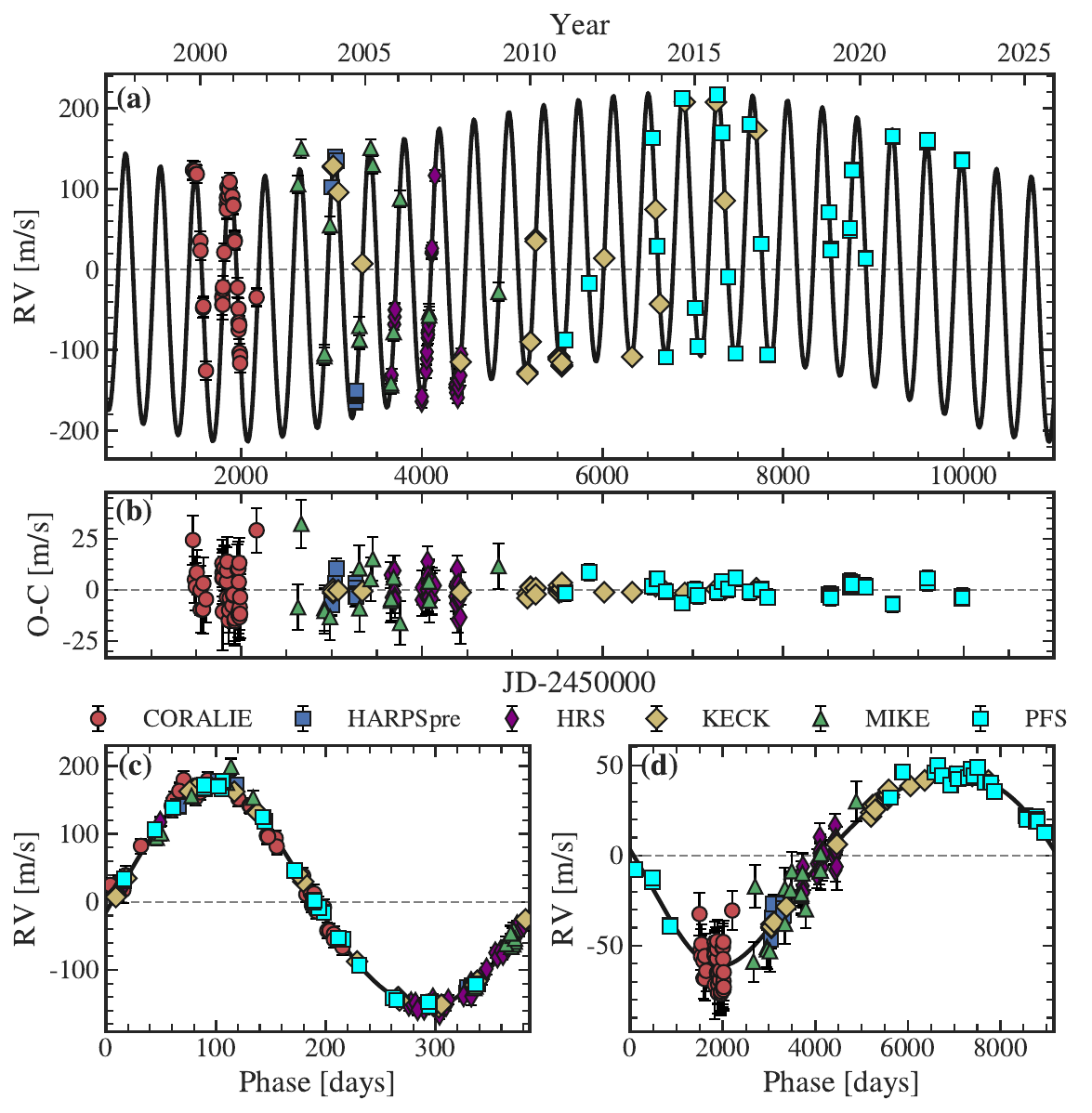}
    \caption{RV+HG23 fits to HD\,28185 RVs from Model 1. Panel (a) shows the best-fit Keplerian orbit (thick black line) to the RV measurements and Panel (b) show their residuals. Panel (c) shows the phase-folded orbit of the inner planet HD\,28185\,b, with the signal of the outer planet HD\,28185\,c being subtracted. Likewise, Panel (d) shows the phase-folded orbit of HD\,28185\,c after correcting the signal of HD\,28185\,b.}
    \label{fig:HD28185_rv}
\end{figure*}

First, F22 and V24 use different RV datasets. Unlike V24, F22 omits the CORALIE \citep{udry00} data collected before JD2452500, leading to a baseline about 700\,d shorter than V24. The CORALIE data show a negative RV trend before a turnaround around JD2452000 (see panel (a) of Fig. \ref{fig:HD28185_rv}). Because this turning point provides a strong constraint on the orbital period in V24,  both F22 and \cite{rosenthal21} determine a longer orbital period and higher mass for HD 28185 c without long enough RV baseline to resolve this turning point. In this case, Hipparcos-Gaia astrometry offers limited constraint on the period due to mass-period degeneracy in the astrometric signal.

Second, the presence of a year-long inner planet complicates the astrometric fit. As noted by V24, the inner companion induces an astrometric signal of at least 0.14 mas, comparable to the outer companion’s influence. Additionally, the 9000-day orbital period derived for HD 28185 c by V24 closely aligns with the 24.75-year Hipparcos-Gaia baseline, which could significantly diminish the observed astrometric signal if this solution is accurate.

V24 argue that the inner planet’s orbital period, being similar to Earth’s annual motion, would cause its astrometric effect to average out over the Gaia and Hipparcos observation spans. While this may hold for the differences in proper motion and position between Hipparcos and Gaia, it does not apply to parallax. The discrepancy between the Hipparcos parallax ($23.62 \pm 0.87$\,mas) and Gaia DR3 parallax ($25.48 \pm 0.02$\,mas)—a difference of about 2$\sigma$—suggests a bias possibly introduced by the inner companion. Notably, Gaia DR2 gives a parallax of $25.36 \pm 0.04$\,mas, differing from the Gaia DR3 value by about 3$\sigma$.

We apply the F23 method, which accounts for the astrometric signal of the inner companion, to analyze data from Gaia DR2, DR3, and Hipparcos (collectively referred to as ``HG23''). We use RV data from CORALIE along with RV data from the Carnegie Planet Finder Spectrograph (PFS; \citealt{crane10}), the High-Resolution Spectrograph (HRS; \citealt{tull98}) on the Hobby-Eberly Telescope (HET; \citealt{ramsey98}), the Magellan Inamori Kyocera Echelle (MIKE) spectrograph \citep{bernstein03}, and the High Accuracy Radial velocity Planet Searcher (HARPS; \citealt{pepe00}). 

\begin{figure*}
    \centering
	\includegraphics[width=0.9\textwidth]{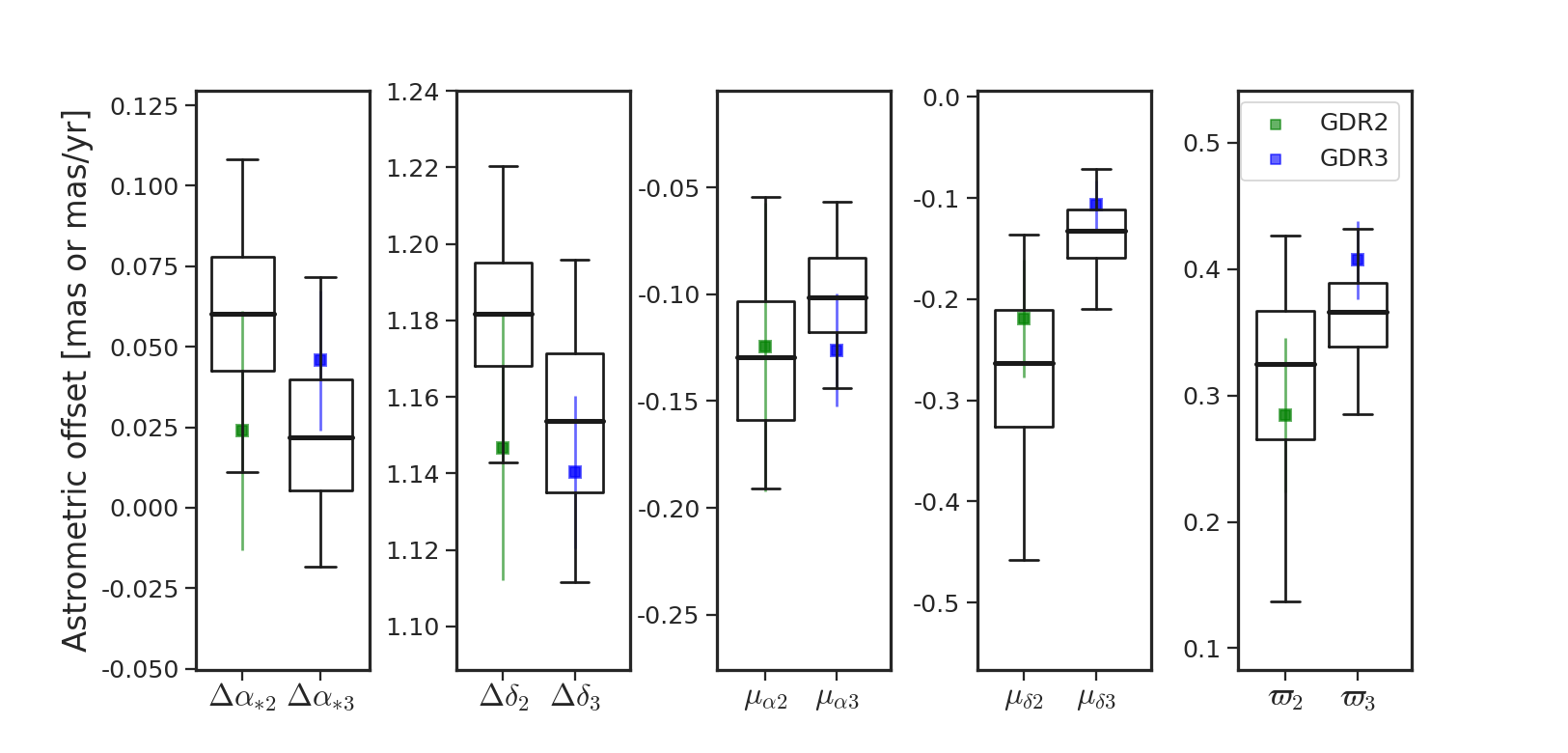}
    \caption{Comparing the five-parameter astrometry of the model 1 prediction to GDR2 and GDR3 astrometry. The barycentric motion of the HD\,28185 system has been subtracted for both catalog Gaia data (square) and the predictions (boxplot). The inner thick line, edge of box, and whisker respectively denote the median, $1\,\sigma$ uncertainty and $3\,\sigma$ uncertainty. The uncertainty is the product of the observed uncertainty and the error inflation factor.
    The subscripts of the label of the x-axis correspond to the Gaia release number.}
    \label{fig:HD28185_boxplot}
\end{figure*}

\begin{figure*}
    \centering
	\includegraphics[width=0.9\textwidth]{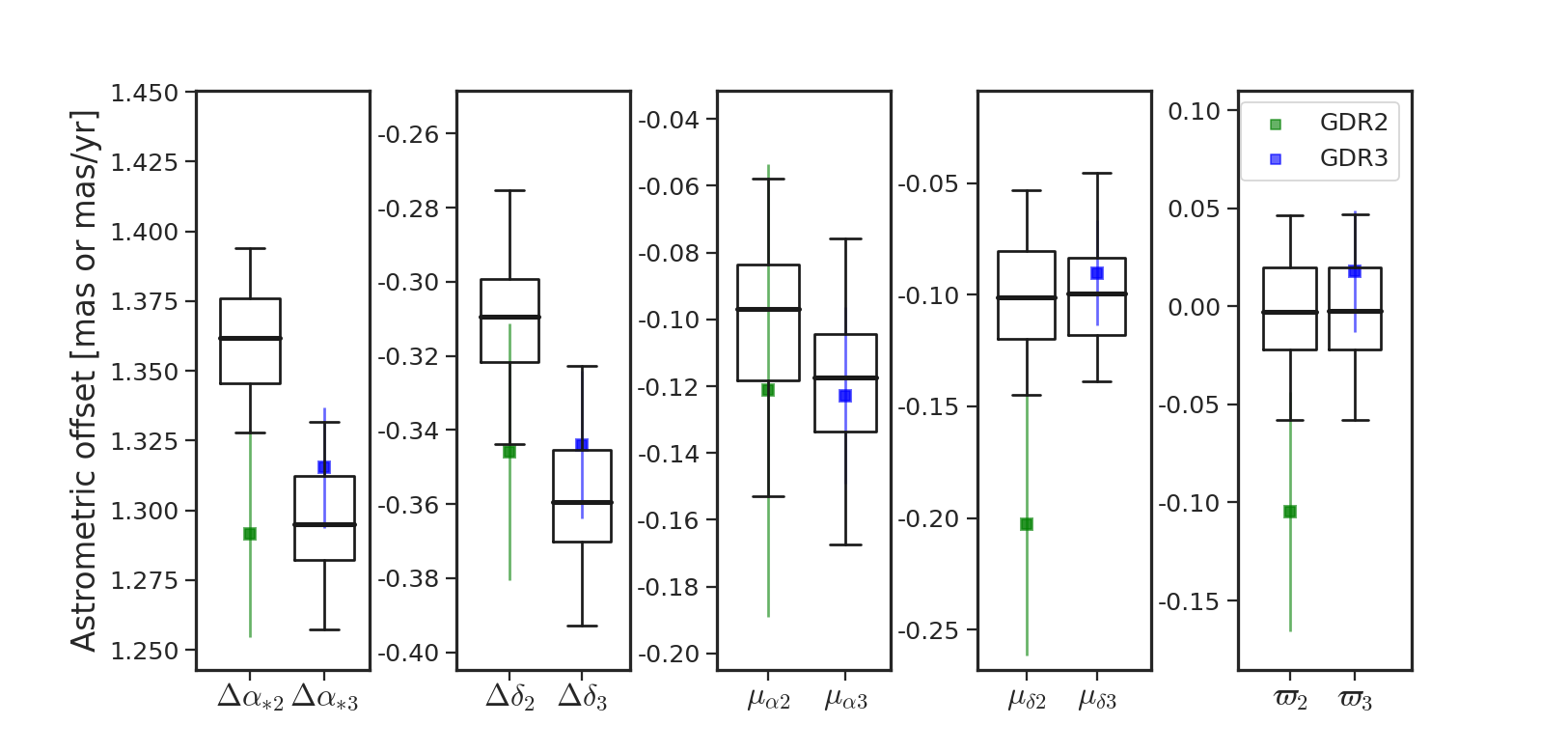}
    \caption{Comparing the five-parameter astrometry of the model 2 prediction to GDR2 and GDR3 astrometry. Symbols are the same as Fig. \ref{fig:HD28185_boxplot}.}
    \label{fig:HD28185_boxplot2}
\end{figure*}

Using the F23 method, we analyze the data and present two solutions: one including the inner companion’s astrometric signal (Model 1) and one excluding it (Model 2). Additionally, we include V24's solution (their Model 1) in Table~\ref{tab:HD28185}. The corresponding fitting results for Model 1 are displayed in Fig.~\ref{fig:HD28185_rv} and Fig.~\ref{fig:HD28185_boxplot}, while the astrometric fitting results for Model 2 are shown in Fig.~\ref{fig:HD28185_boxplot2}. As shown in Table \ref{tab:HD28185}, using the same RV data and excluding the inner companion's astrometric signal from the solution, Model 2 aligns with V24's results. The 1.5-$\sigma$ discrepancy in $\omega_c$ between Model 2 and V24 likely arises from their differing approaches to systematic bias in astrometric catalogs.

As shown in Figs. \ref{fig:HD28185_boxplot}, the inner companion induces a significant parallax offset of approximately 0.4\,mas, consistent with the observed discrepancies among Hipparcos, Gaia DR2, and DR3. This contribution from the inner companion to the total astrometric signal results in an improvement in the model’s fit, with an increase in the log-likelihood by 8.4. This corresponds to a decrease in the Bayesian Information Criterion (BIC) by 12, assuming two additional parameters ($I_b$ and $\Omega_b$) compared to the model that excludes the inner companion’s influence. The improvement in orbital fit is also evident in the comparison between models 1 and 2 applied to the Gaia DR2 and DR3 data (see Figs. \ref{fig:HD28185_boxplot} and \ref{fig:HD28185_boxplot2}).

Additionally, including the astrometric signal from the inner companion increases the uncertainties in the estimates of $I_c$ and $\Omega_c$ (see Fig. \ref{fig:corner_HD28185}). This is likely because some astrometric variation is attributed to the inner companion, reducing the signal-to-noise ratio for the outer companion. Despite these differences, models 1 and 2 produce nearly identical solutions for the other shared orbital parameters (see Fig. \ref{fig:corner_HD28185}).
Our two solutions for the outer companion are consistent with those presented in V24. Since V24 and \texttt{orvara} follow the first convention, while we use the astrometric convention defined in \cite{feng19c}, our value of $\Omega$ differs from theirs by 180\,\text{deg}.

While the use of both DR2 and DR3 does not fully resolve the $I$ and $\pi-I$ degeneracy for the outer companion with a 25-year orbit, it provides a unique inclination solution for the inner companion, as shown in Table \ref{tab:HD28185} and Fig. \ref{fig:corner_HD28185}. This demonstrates the importance of using both Gaia DR2 (J2015.5) and DR3 (J2016.0) in uniquely determining the inclination of orbits with periods comparable to, or not much longer than, the half-year difference between the DR2 and DR3 reference epochs.

Based on the analyses above, we conclude that the discrepancy between the solutions for HD 28185 c reported by F22 and V24 is primarily due to differences in the RV data. A limited RV baseline can result in degeneracies between a companion's mass and orbital period. These degeneracies may be mitigated by extending the RV baseline or incorporating relative astrometry data, as demonstrated in \citet{philipot23a,mesa23,elmorsy24}. This issue will be explored in greater detail in the following sections.

\begin{table*}
\renewcommand{\arraystretch}{1.5}
\centering
\caption{Parameters for HD\,28185 system. Model 1 includes the astrometric contribution from the inner planet HD\,28185\,b, while Model 2 ignores it. Values in brackets represent alternative solutions.}\label{tab:HD28185}
\begin{tabular*}{\textwidth}{@{}@{\extracolsep{\fill}}lcccccc@{}}
\hline \hline 
Parameter & \multicolumn{2}{c}{Model 1} & \multicolumn{2}{c}{Model 2} & \multicolumn{2}{c}{V24}\\
 & HD\,28185\,b & HD\,28185\,c & HD\,28185\,b & HD\,28185\,c & HD\,28185\,b & HD\,28185\,c\\
\hline
\textbf{Fitted Parameter}&&&&\\
Orbital period $P$ (day)&${385.858}_{-0.055}^{+0.054}$&${9229}_{-222}^{+331}$ &${385.859}_{-0.052}^{+0.053}$&${9236}_{-222}^{+319}$&${385.92}_{-0.07}^{+0.06}$&${9090}_{-390}^{+460}$\\
RV semi-amplitude $K$ (m s$^{-1}$) &${163.67}_{-0.71}^{+0.74}$&${52.8}_{-2.4}^{+2.8}$ &${163.67}_{-0.73}^{+0.71}$&${52.7}_{-2.3}^{+3.0}$&${164.8}_{-0.09}^{+0.09}$&${53.3}_{-4.7}^{+5.1}$\\
Eccentricity $e$&${0.0634}_{-0.0030}^{+0.0030}$&${0.140}_{-0.026}^{+0.024}$ &${0.0634}_{-0.0030}^{+0.0029}$&${0.139}_{-0.024}^{+0.023}$&${0.063}_{-0.004}^{+0.004}$&${0.15}_{-0.04}^{+0.04}$\\
Argument of periapsis${\rm ^a}$ $\omega$ (deg)&${358.4}_{-2.7}^{+2.8}$&${148.4}_{-5.1}^{+4.5}$ &${358.4}_{-2.7}^{+2.7}$&${148.2}_{-5.3}^{+4.5}$&${355.1}_{-3.9}^{+3.9}$&${162}_{-8}^{+8}$\\
Mean anomaly at JD2451463 $M_0$ (deg)&${341.3}_{-2.9}^{+2.8}$&${6.7}_{-4.9}^{+10}$ &${341.3}_{-2.9}^{+2.8}$&${6.7}_{-4.8}^{+10}$&---&---\\
Inclination $I$ (deg)&${156.5}_{-9.5}^{+6.1}$&${73.0}_{-8.4}^{+10}$ (${109.5}_{-12}^{+9.1}$) &---&${64.9}_{-5.9}^{+7.2}$ (${111.8}_{-10}^{+6.8}$)&---&${66}_{-9}^{+11}$ (${114}_{-11}^{+9}$)\\
Longitude of ascending node${\rm ^b}$ $\Omega$ (deg)&${28}_{-22}^{+29}$&${107}_{-22}^{+19}$ (${40}_{-27}^{+28}$) &---&${93}_{-14}^{+13}$ (${-5}_{-13}^{+17}$)&---&${271}_{-21}^{+15}$ (${178}_{-14}^{+18}$)\\\hline
\textbf{Derived Parameter}&&&&\\
Orbital period $P$ (yr)&${1.05642}_{-0.00015}^{+0.00015}$&${25.27}_{-0.61}^{+0.91}$ &${1.05643}_{-0.00014}^{+0.00014}$&${25.29}_{-0.61}^{+0.87}$&${1.0566}_{-0.0002}^{+0.0002}$&${24.9}_{-1.1}^{+1.3}$\\
Semi-major axis $a$ (au)&${1.0282}_{-0.0064}^{+0.0063}$&${8.54}_{-0.14}^{+0.21}$ &${1.0285}_{-0.0064}^{+0.0063}$& ${8.54}_{-0.15}^{+0.20}$&${1.034}_{-0.006}^{+0.006}$& ${8.50}_{-0.26}^{+0.29}$\\
Companion mass${\rm ^c}$ $m$ ($M_{\rm Jup}$)&${13.3}_{-3.9}^{+5.0}$&${5.68}_{-0.36}^{+0.44}$ &---&${5.90}_{-0.38}^{+0.41}$&---&${6.0}_{-0.6}^{+0.6}$\\
Periapsis epoch $T_{\rm p}-2450000$ (JD)&${1868.8}_{-3.0}^{+3.1}$&${10500}_{-162}^{+169}$ &${1868.7}_{-3.0}^{+3.1}$&${10502}_{-162}^{+169}$&${1870.2}_{-4.5}^{+4.5}$&${10790}_{-280}^{+350}$\\
\hline
\textbf{Barycentric Offset}&&&&\\
$\alpha*$ offset $\Delta \alpha*$ (mas)& \multicolumn{2}{c}{${1.07}_{-0.22}^{+0.16}$ (${0.65}_{-0.52}^{+0.36}$)}&\multicolumn{2}{c}{${1.28}_{-0.11}^{+0.12}$ (${-0.22}_{-0.30}^{+0.41}$)}&\multicolumn{2}{c}{---}\\
$\delta$ offset $\Delta \delta$ (mas)&\multicolumn{2}{c}{${-0.50}_{-0.37}^{+0.48}$ (${0.95}_{-0.45}^{+0.23}$)}&\multicolumn{2}{c}{${-0.19}_{-0.31}^{+0.35}$ (${1.23}_{-0.11}^{+0.13}$)}&\multicolumn{2}{c}{---}\\
$\mu_{\alpha*}$ offset $\Delta \mu_{\alpha*}$ (mas\,yr$^{-1}$)&\multicolumn{2}{c}{${-0.107}_{-0.024}^{+0.022}$}&\multicolumn{2}{c}{${-0.103}_{-0.023}^{+0.024}$}&\multicolumn{2}{c}{---}\\
$\mu_\delta$ offset $\Delta \mu_\delta$ (mas\,yr$^{-1}$)&\multicolumn{2}{c}{${-0.098}_{-0.017}^{+0.017}$}&\multicolumn{2}{c}{${-0.094}_{-0.018}^{+0.018}$}&\multicolumn{2}{c}{---}\\
$\varpi$ offset $\Delta \varpi$ (mas)&\multicolumn{2}{c}{${0.280}_{-0.081}^{+0.090}$}&\multicolumn{2}{c}{${0.017}_{-0.021}^{+0.021}$}&\multicolumn{2}{c}{---}\\\hline
\textbf{Instrumental Parameter}&&&&\\
RV offset for CORALIE (m\,s$^{-1}$)&\multicolumn{2}{c}{${50306.4}_{-3.4}^{+3.4}$}&\multicolumn{2}{c}{${50306.3}_{-3.3}^{+3.6}$}&\multicolumn{2}{c}{${50305.9}_{-8.2}^{+8.2}$}\\
RV offset for HARPSpre (m\,s$^{-1}$)&\multicolumn{2}{c}{${71.8}_{-2.6}^{+1.7}$}&\multicolumn{2}{c}{${71.8}_{-2.5}^{+1.6}$}&\multicolumn{2}{c}{${75.6}_{-4.3}^{+4.0}$}\\
RV offset for HRS (m\,s$^{-1}$)&\multicolumn{2}{c}{${93.7}_{-2.8}^{+2.2}$}&\multicolumn{2}{c}{${93.7}_{-2.8}^{+2.2}$}&\multicolumn{2}{c}{${95.1}_{-4.4}^{+4.3}$}\\
RV offset for PFS (m\,s$^{-1}$)&\multicolumn{2}{c}{${-48.6}_{-2.6}^{+1.9}$}&\multicolumn{2}{c}{${-48.7}_{-2.6}^{+2.0}$}&\multicolumn{2}{c}{${-31.5}_{-4.2}^{+4.0}$}\\
RV offset for KECK (m\,s$^{-1}$)&\multicolumn{2}{c}{${89.4}_{-2.6}^{+1.9}$}&\multicolumn{2}{c}{${89.4}_{-2.6}^{+2.0}$}&\multicolumn{2}{c}{---}\\
RV offset for MIKE (m\,s$^{-1}$)&\multicolumn{2}{c}{${53.1}_{-2.6}^{+1.7}$}&\multicolumn{2}{c}{${53.1}_{-2.5}^{+1.7}$}&\multicolumn{2}{c}{${55.3}_{-4.9}^{+4.6}$}\\
RV jitter for CORALIE (m\,s$^{-1}$)&\multicolumn{2}{c}{${9.7}_{-1.7}^{+2.0}$}&\multicolumn{2}{c}{${9.6}_{-1.7}^{+2.0}$}&\multicolumn{2}{c}{${9.0}_{-1.8}^{+2.1}$}\\
RV jitter for HARPSpre (m\,s$^{-1}$)&\multicolumn{2}{c}{${5.7}_{-1.3}^{+1.9}$}&\multicolumn{2}{c}{${5.7}_{-1.3}^{+2.1}$}&\multicolumn{2}{c}{${6.0}_{-1.6}^{+2.3}$}\\
RV jitter for HRS (m\,s$^{-1}$)&\multicolumn{2}{c}{${1.6}_{-1.2}^{+1.6}$}&\multicolumn{2}{c}{${1.6}_{-1.1}^{+1.6}$}&\multicolumn{2}{c}{${1.8}_{-1.2}^{+1.7}$}\\
RV jitter for PFS (m\,s$^{-1}$)&\multicolumn{2}{c}{${3.95}_{-0.56}^{+0.69}$}&\multicolumn{2}{c}{${3.95}_{-0.57}^{+0.69}$}&\multicolumn{2}{c}{${4.1}_{-0.7}^{+1.0}$}\\
RV jitter for KECK (m\,s$^{-1}$)&\multicolumn{2}{c}{${1.21}_{-0.36}^{+0.37}$}&\multicolumn{2}{c}{${1.21}_{-0.35}^{+0.38}$}&\multicolumn{2}{c}{---}\\
RV jitter for MIKE (m\,s$^{-1}$)&\multicolumn{2}{c}{${12.3}_{-2.7}^{+3.3}$}&\multicolumn{2}{c}{${12.3}_{-2.6}^{+3.3}$}&\multicolumn{2}{c}{${12.2}_{-2.6}^{+3.4}$}\\
\hline
Jitter for hipparcos $J^{\rm hip}$ (mas)&\multicolumn{2}{c}{${2.01}_{-0.54}^{+0.52}$}&\multicolumn{2}{c}{${2.03}_{-0.51}^{+0.51}$}&\multicolumn{2}{c}{---}\\
Error inflation factor $S^{\rm gaia}$&\multicolumn{2}{c}{${1.079}_{-0.054}^{+0.076}$}&\multicolumn{2}{c}{${1.150}_{-0.076}^{+0.081}$}&\multicolumn{2}{c}{---}\\
\hline 
\multicolumn{7}{l}{$^{\rm a}$ The argument of periastron of the stellar reflex motion, differing by $\pi$ with planetary orbit, i.e., $\omega_{\rm p}=\omega+\pi$.} \\
\multicolumn{7}{l}{$^{\rm b}$ The values of $\Omega$ should be increased by 180\,\text{deg} for comparison with those reported by V24. }\\
\multicolumn{7}{l}{$^{\rm c}$ The stellar mass of $0.974\pm0.018\,M_{\odot}$ is adopted from V24 and is assigned a Gaussian prior.}\\
\end{tabular*}
\end{table*}

\section{Other Discrepancies in F22's Orbital Solutions}\label{sec:discrepancy}
Aside from our detailed worked example of HD28185 in the previous section, we also discuss several other targets illuminated by V24 as showing discrepancies between F22 and other studies. We divide the causes of discrepancies into four categories, including different conventions (section \ref{sec:convention}), insufficient posterior sampling (section \ref{sec:sampling}), inner companions (section \ref{sec:inner}), and limited RV coverage (section \ref{sec:coverage}). We will also address the mass-period degeneracy in section \ref{sec:degeneracy} and explore the anisotropic inclination distribution in section \ref{sec:inclination}. Since both \texttt{orvara} and F23 adopt the same parallel-tempering Markov Chain Monte Carlo sampler \texttt{ptemcee}\citep{Vousden2016}, we set 30 temperatures, 100 walkers, and 50,000 steps (80,000 for 14\,Her) per chain to generate posterior distributions for all the fitting parameters, with the first 25,000 steps being discarded as burn-in. The priors can be found in \citet{brandt21c} and \citet{feng23}, respectively. By default, the HGCA version for \texttt{orvara} is chosen to EDR3. The RV datasets used here for HD\,38529, 14\,Her and GJ\,229 are the same as \citet{xuan20}, \citet{bardalez21} and \citet{brandt21d}, respectively.

\subsection{Different conventions}\label{sec:convention}
HD 38529 hosts at least two companions, HD 38529 has been studied by \cite{benedict10}, \cite{xuan20}, and F22. Using HST FGS data, \cite{benedict10} derived $I_c=48.8\pm4.0$\,deg and $\Omega_c=37.8\pm8.2$\,deg for the outer companion. In contrast, \cite{xuan20} report $I_c=135_{-14}^{+8}$\,deg and $\Omega_c=217_{-19}^{+15}$\,deg based on Hipparcos and Gaia data. Notably, HGCA-based methods often yield two solutions for $I$ and $\Omega$, mirrored around $I_c=90$\,deg (see fig. 3 of V24). V24 suggest adjusting $(I_c, \Omega_c)$ given by \cite{benedict10} to align with \cite{xuan20} considering their use of different conventions. Without this adjustment, $\Omega_c$ determined by \cite{benedict10} is consistent with F22's $\Omega_c=37.8_{-14.9}^{+16.2}$\,deg, and  $I_c$ estimated by \cite{xuan20} differs from F22’s $104.6_{-8.7}^{+6.4}$\,deg by $2\sigma$. 

\begin{figure*}
    \centering
	\includegraphics[width=0.9\textwidth]{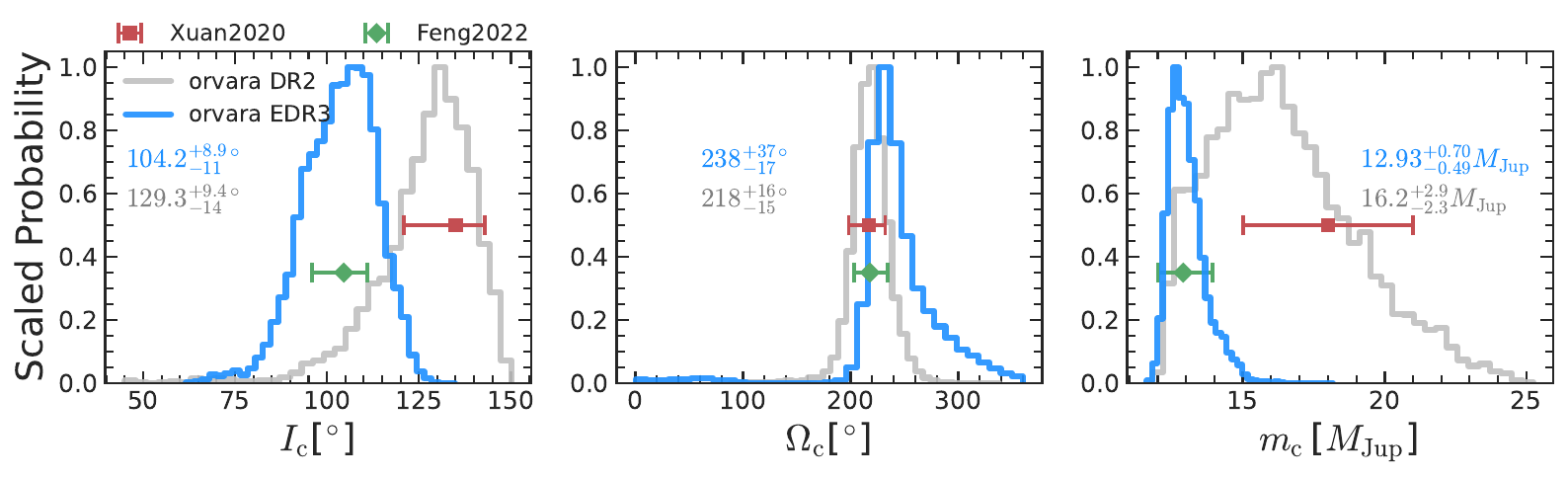}
    \caption{Marginalized posteriors of inclination $I$, longitude of ascending node $\Omega$ and mass $m$  for HD\,38529\,c using \texttt{orvara} HGCA DR2 (gray; \citealt{brandt18}) and EDR3 (blue; \citealt{brandt21}), respectively. Their median and $1\sigma$ confidence intervals are shown with the corresponding colors in each panel. The y-axis is scaled between 0 and 1. 
    The results of \citet{xuan20} and \citet{feng22} are respectively denoted with the red square and the green diamond. It should be note that the value of $\Omega$ reported by \citet{feng22} is advanced by 180\,deg to match the convention of \texttt{orvara}. It is clear that the main discrepancy between two studies stems from the use of different Gaia data.}
    \label{fig:HD38529_GDR_comp}
\end{figure*}

The \(\sim 2\sigma\) discrepancy between the inclinations given in F22 and \citet{xuan20} is likely due to the former using Gaia EDR3, while the latter used Gaia DR2. This is demonstrated by comparing the solutions from \cite{xuan20} and F22 with the \texttt{orvara} solutions derived from Gaia DR2 and EDR3 data in Fig. \ref{fig:HD38529_GDR_comp}. Additionally, \citet{xuan20} determined a disk inclination of $71_{-7}^{+10}$\,deg or equivalently $109_{-10}^{+7}$\,deg, which agrees with F22 but differs from the planet c inclination reported by \citet{xuan20} by \(\sim 2\sigma\). Given that the HST-derived inclination may be biased toward face-on solutions \citep{benedict22}, the inclination uncertainty given by \cite{benedict10} may be underestimated.

Considering that \texttt{orvara} uses the first convention defined in \cite{feng19c} and F22 use the third convention (or astrometric convention) defined in \cite{feng19c}, the longitude of ascending nodes given by F22 and \texttt{orvara} differ by 180\,\text{deg}. This difference in convention explains the so-called 11$\sigma$ discrepancy in $\Omega$ reported by F22 and \cite{xuan20}.

\subsection{Insufficient posterior sampling}\label{sec:sampling}
\begin{figure*}
    \centering
	\includegraphics[width=\textwidth]{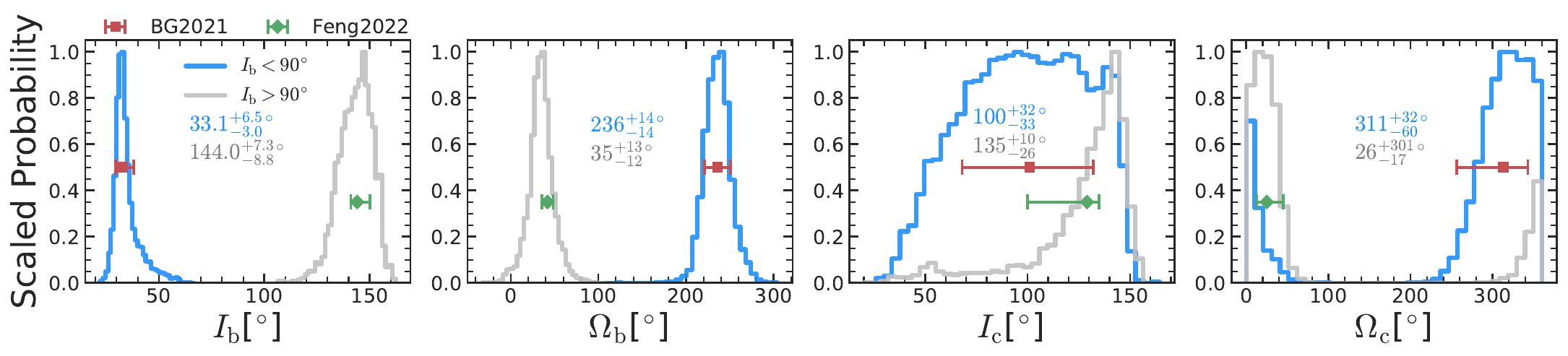}
    \caption{Marginalized posteriors of inclination $I$ and longitude of ascending node $\Omega$ for 14\,Her\,b and c using \texttt{orvara} (EDR3 version). The posteriors are separately displayed based on the inclination of 14\,Her\,b, i.e., $I_{\rm b}<90^{\circ}$ (blue) and $I_{\rm b}>90^{\circ}$ (gray). The results of \citet{bardalez21} (BG2021) and \citet{feng22} are also shown with the red square and the green diamond. The same adjustment for $\Omega$ of \citet{feng22} has been done for fair comparison. It seems that both studies only found one of the two solutions. }
    \label{fig:corner_comp_14Her}
\end{figure*}

14 Her, with two wide companions, has been studied by \cite{bardalez21}, who derived inclinations of $32.7_{-3.2}^{+5.3}$\,deg and $101_{-33}^{+31}$\,deg using HGCA data analyzed with \texttt{orvara}. \cite{benedict23} found inclinations of $35.7\pm3.2$\,deg and $82\pm14$\,deg for 14 Her b and c, while F22 reported $144_{-3}^{+6}$\,deg and $120_{-29}^{+6}$\,deg. Missing the alternative inclination solution for 14 Her b from \cite{bardalez21} (147.3\,deg) creates an apparent discrepancy, yet this solution is consistent with F22’s value. In Fig.~\ref{fig:corner_comp_14Her}, we compare the solutions from F22 and \cite{bardalez21} with the \texttt{orvara} solutions rederived in this work. It is evident that both studies missed the alternative solution, and their inclination estimates are mirrored around 90$^\circ$. Such insufficient posterior sampling for inclination occurs in many \texttt{orvara}-based studies and F22. For example, the solution \(I \sim 45^\circ\) is only partially resolved by \cite{xuan20} using parallel-tempered (PT) Markov chain Monte Carlo (MCMC) (see the right panel of their figure 2) and is not presented as an alternative inclination solution. In our analyses, we employ \texttt{ptemcee} to sample the posterior and use the Gelman-Rubin statistic \(\hat{R} < 1.1\) \citep{gelman92} as a convergence criterion. However, even with \(\hat{R} < 1.1\), chains may fail to sample other modes of a multi-modal posterior distribution. Therefore, future studies should consider running multiple chains with diverse initializations and utilize a broader range of posterior sampling techniques to ensure thorough exploration of the posterior distribution.

\subsection{Multiple companions}\label{sec:inner}
GJ 229 B is a well-known brown dwarf analyzed by \cite{brandt20}, \cite{brandt21d}, and F22. V24 noted that F22’s estimated mass of $60.4_{-2.4}^{+2.3}$\mj is lower than the 71.4$\pm$0.6\mj given by \cite{brandt21d}, likely because F22 accounted for RV variations from two inner planets with periods of approximately 120 and 520 days \citep{feng20a}, while other studies did not. To demonstrate this, we constrain the mass and orbit of GJ 229 B using the same data as in \cite{brandt21d}, with the corner plot shown in Fig.~\ref{fig:corner_GJ229}. Our best-fit mass of $71.73_{-0.56}^{+0.55}$\mj aligns perfectly with the value reported in \cite{brandt21d}. However, discrepancies in other parameters, such as eccentricity, likely arise from our treatment of HARPS RV data before and after the fiber change as two independent data sets, whereas \cite{brandt21d} combines them into a single set. The mass estimation for GJ 229 B (a binary system comprising GJ 229 Ba and Bb, as reported by \citet{xuan24}), derived without including potential inner companions, aligns with the dynamical mass inferred by \cite{xuan24} using the data from the Very Large Telescope (VLT) interferometer in GRAVITY Wide mode and the CRyogenic InfraRed Echelle Spectrograph Upgrade Project (CRIRES+). This suggests that the existence of inner companions cannot be confirmed by the current astrometric and RV data.

HD 62364 hosts at least one wide companion, HD 62364 has been examined by F22 and others \citep{xiao23,frensch23,philipot23a}. While F22 identified two companions using HARPS data before 2019, subsequent studies with additional HARPS data found only one. By examining the one-companion and two-companion solutions from F22, we identified that the two solutions used different RV datasets. The one-companion solution included 14 additional RVs compared to the two-companion solution. Due to an oversight in the dataset change, F22 incorrectly calculated the logarithmic Bayes Factor (lnBF) using the same number of data points, which led to the erroneous model selection. While such errors are rare, they can be avoided through careful management of databases and orbital solutions, particularly when analyzing large samples of targets.

By reanalyzing the updated RV data, we find that the one-companion solution (Model 1) is strongly favored over the two-companion solution (Model 2), with a \(\ln\)BF\(_{21}\) of approximately \(-8\). To compare with the results for HD 62364 b from \cite{xiao23}, \cite{frensch23}, and \cite{philipot23b}, we reanalyzed the updated HARPS RV data and Hipparcos-Gaia astrometric data. The best-fit parameters are presented in Table \ref{tab:pars1}, and the corresponding posterior distributions are shown in Fig. \ref{fig:corner_HD62364}. As seen from Fig. \ref{fig:corner_HD62364}, our results are consistent with previous solutions but with higher precision due to use of multiple Gaia DRs.

Therefore, the investigations of these discrepancies highlight the critical role of additional data, accurate model inference, and the determination of the correct number of planetary signals in deriving consistent orbital solutions.

\subsection{Limited RV coverage and lack of relative astrometry}\label{sec:coverage}
HD 211847, GJ 680 and HD 111031 host low-mass stellar companions identified through imaging \cite{moutou17,ward15,gonzales20,dalba21}. F22 initially classified them as brown dwarfs due to RV-only data, though recent studies including relative astrometry identified them as low-mass stars \citep{philipot23a,philipot23b}. Without relative astrometry from imaging data, \cite{philipot23a} acknowledge that they cannot obtain the correct solution using only RV and absolute astrometry, due to the limited RV time span. The oversight of the relative astrometry data for these three targets by F22 is likely due to the challenges of comprehensively searching the literature for such a large sample of targets, combined with their primary emphasis on analyzing combined RV and astrometry data.

Using relative and absolute astrometry alongside RV data from \cite{philipot23a,philipot23b}, we derive orbital solutions for HD 211847 B, GJ 680 B, and HD 111031 B using both the F23 and \texttt{orvara} methods (Table \ref{tab:pars2}; Figs. \ref{fig:corner_HD211847}–\ref{fig:corner_HD111031}). For HD 211847, we added 51 HARPS RVs from the ESO archive to the RV dataset used in \cite{philipot23a,philipot23b}.

The solutions for HD 211847 B derived using F23 and \texttt{orvara} are consistent with each other and with the solution provided by \cite{philipot23a} within 2$\sigma$. For GJ 680 B, while the F23 and \texttt{orvara} solutions are consistent, the F23 solution exhibits significantly smaller uncertainty, likely due to the incorporation of Gaia DR2 data. Similarly, the F23 and \texttt{orvara} solutions for HD 111031 B are consistent within 1$\sigma$, as illustrated in Fig.~\ref{fig:corner_HD111031}. This further demonstrates that the discrepancies between F22 and other studies regarding these three targets stem from F22's oversight of relative astrometry data, particularly when RV coverage is limited, rather than from methodological differences.

\subsection{Mass-period degeneracy}\label{sec:degeneracy}
A key reason behind the discrepancies is that RV trends induced by long-period companions mainly constrain the host star’s acceleration ($g$) due to the companion. For a companion of mass $m$ on a circular orbit with semi-major axis $a$ and inclination $I$, the RV acceleration is $g\propto m\sin{I}/a^2$. Thus, RV trends primarily constrain $m\sin{I}/a^2$, with astrometry from Hipparcos-Gaia contributing information on $I$ and the longitude of the ascending node $\Omega$. This limitation explains why values of $m/a^2$ reported by F22 align with other studies, despite different $(m, a)$ values. However, longer RV coverage or relative astrometry is essential to constrain eccentric orbits and break the $m-a$ (or mass-period $m-P$) degeneracy. 
Even if relative astrometry is not available from direct imaging data, the null results from such imaging can still constrain the potential solutions for the companions \citep{mawet19}.

Our detailed comparison in Section \ref{sec:comparison} confirms that the F19 method used in F22 is equivalent to \texttt{orvara}. V24 suggested that the discrepancies in solutions are period-dependent. As discussed, in most cases, the root cause of discrepancies for long-period companions is the partial RV coverage rather than methodology.  This ``evolution'' of solutions for long-period companions, as detailed in Section \ref{sec:epsIndA}, largely arises from short RV baselines.

\subsection{Inclination distribution}\label{sec:inclination}
V24 also argued that inclination discrepancies for long-period
companions result from the non-isotropic inclination distribution
observed by \cite{benedict23}. However, as shown in \cite{benedict23},
neither the HST exoplanet sample nor F22’s sample follows the
$\cos{I}$-uniform distribution seen in the 6th Visual Binary Star
Catalog \citep{hartkopf01}. This is because RV data, sensitive to
$m\sin{I}$, favors high-inclination orbits, while astrometric signals
inversely relate to inclination. Consequently, the inclination
distribution for RV and astrometric detections differs from that of
binaries with more prominent signals unaffected by detection limits.

\section{Evolution of solutions for eps Ind A b}\label{sec:epsIndA}
Following initial indications of a potential wide-orbit companion in
the RV data \citep{endl02}, continued efforts have been made to monitor and image
this companion, culminating in the first successful image captured by \cite{matthews24} using JWST/MIRI. The imaging of this system was guided by
combined analyses of RV and Hipparcos-Gaia data
\citep{feng19b,philipot23a,feng23}. The solutions and data from these
studies are summarized in Table \ref{tab:epsIndA}. 

In this work, we use the RV data collected by the Coudé Echelle Spectrometer Long Camera (LC) and the Very Long Camera (VLC) \citep{Zechmeister2013}, the ESO UV-visual echelle spectrograph (UVES; \citealt{dekker00}), and HARPS before (HARPSpre) and after (HARPSpost) fibre change, and HARPS during covid pandemic (HARPSpost2). The dataset of HARPSpost2 were released by \citet{Barbieri2023} which extends the time baseline to $\sim29$yr. We bin the data each night to eliminate the high-frequency signal. 

\begin{figure*}
    \centering	\includegraphics[width=\textwidth]{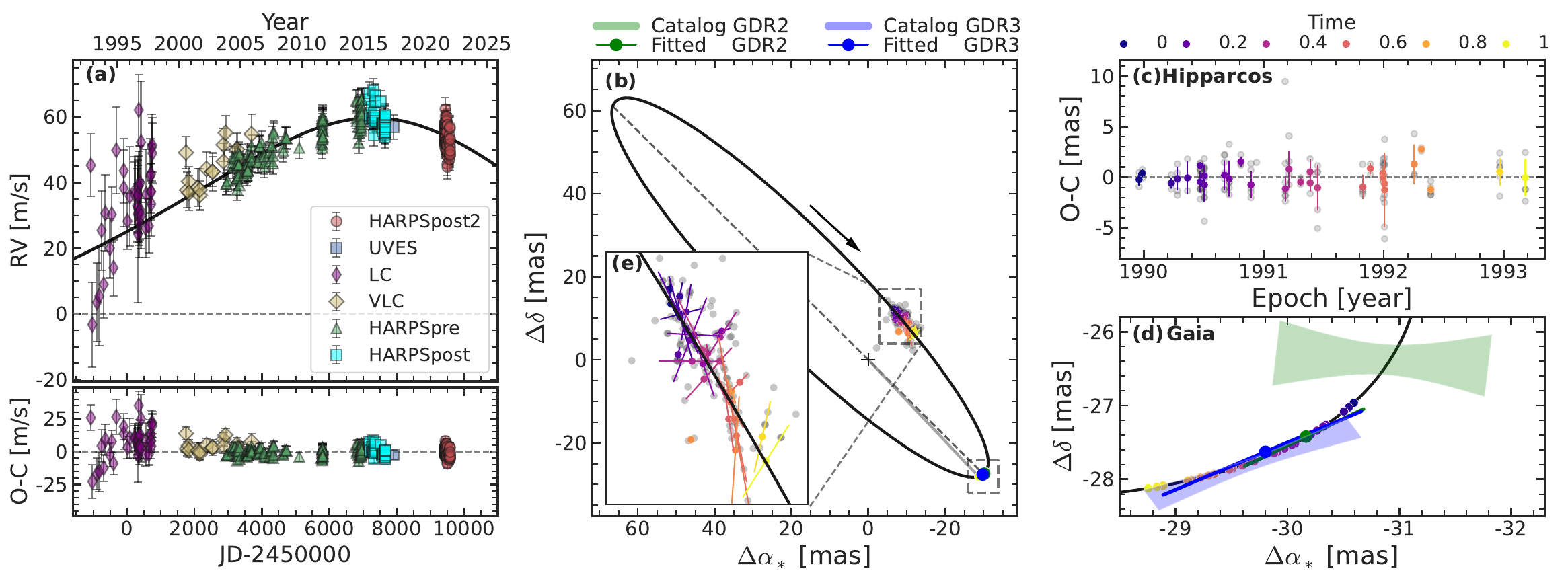}
    \caption{RV+HG23 fits to the RVs, Hipparcos, and Gaia astrometry of eps Ind A (Solution B). Panel (a) shows the RV curve, with the best-fit Keplerian orbit indicated by the thick black line. Residuals (O-C) between the observed RVs and the model are depicted below.
    Panel (b) plots the best-fit astrometric orbit of the star. The black dashed line inside the orbit connects the ascending node and the descending node. The plus symbol denotes the system's barycenter, and the grey line connects it with the periapsis. The post-fit Hipparcos abscissa residuals are projected into the R.A. and decl. axes (grey dots) and have been binned into single points with colors. The brightness of these points gradually increases with observation time (the temporal baseline of each satellite is set to 1). The orientations of the error bars of each point denote the along-scan direction of Hipparcos. The arrow denotes the orientation of the orbital motion. Panel (c) shows the residual (O-C) of Hipparcos abscissa. Panel (c) and (d) are respectively the enlargement of the fitting to Hipparcos and Gaia astrometry. The former magnifies the square region of panel (b) which depicts the best fit to Gaia GOST data and the comparison between best-fit and catalog astrometry (positions and proper motions) at GDR2 and GDR3 reference epochs. The shaded regions represent the uncertainty of catalog positions and proper motions after removing the motion of the system's barycenter. The two segments and their center dots (green and blue) represent the best-fit proper motion and position offsets induced by the planet for Gaia DR2 and DR3.}
    \label{fig:HD209100_rvhg}
\end{figure*}

\begin{figure*}
  \begin{center}
    \includegraphics[scale=0.58]{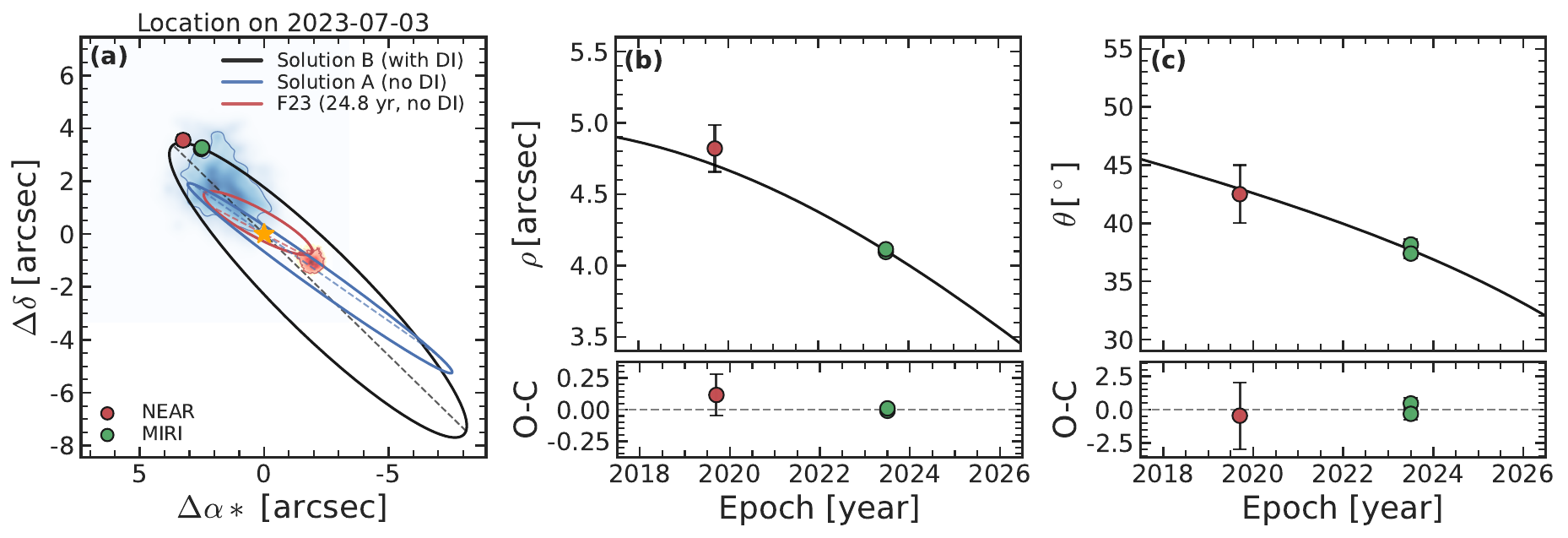}
    \caption{Comparison of the position of $\epsilon$ Ind A b as predicted by F23 with a 24.8-year RV baseline, Solution A, and Solution B, with the observed position from JWST/MIRI. The shade regions of Panel (a) denote the predicted location of the planet ($1\sigma$ uncertainty) on July 3, 2023. Panel (b) and (c) respectively show the best fit of our Solution B to the relative separation ($\rho$) and position angle ($\theta$). }
    \label{fig:HD209100_rel}
\end{center}
\end{figure*}

Solution A and Solution B are derived respectively using the F23 method without
and with direct imaging (DI) data. The DI data were collected by the VISIR/NEAR instrument \citep{Pathak2021,Viswanath2021} at VLT and the MIRI mounted in JWST.
The RV data used in this study largely aligns with that of \cite{matthews24}. However, the perspective acceleration has been subtracted from the LC and VLC RVs in accordance with the recommendations of \cite{janson09}. Comparisons among F23, Solution A, and Solution B are illustrated in Fig. \ref{fig:HD209100_rvhg} and Fig. \ref{fig:HD209100_rel}.

\begin{table}
  \renewcommand{\arraystretch}{1.5}
    \centering
    \caption{Parameters given in various studies for eps Ind A b and the data used by them. }
    \begin{tabular}{l*9{c}}
      \hline\hline
     $\Delta T_{rv}$$^a$&Astrometry&Imaging&Method&$m_b$&$P_b$&$e_b$&$I_b$&$\Omega_b$&Reference\\
      yr&&&&\mj&yr&&deg&deg&\\
      \hline
                    24.8&HG2&---&F19&$3.25_{-0.65}^{+0.39}$&$45.20_{-4.77}^{+5.74}$&$0.26_{-0.03}^{+0.07}$&$64.25_{-6.09}^{+13.80}$&$250.20_{-14.84}^{+14.72}$&\cite{feng19b}\\
                    24.8&HG3&---&\texttt{orvara}-like$^b$&$3.0\pm
                                                      0.1$&$29.9_{-0.6}^{+0.7}$&$0.48\pm0.01$&$91_{-5}^{+4}$&$58\pm 5$&\cite{philipot23a}\\
                    24.8&HG23&---&F23&$2.96_{-0.38}^{+0.41}$&$42.92_{-4.09}^{+6.38}$&$0.26\pm0.04$&$84.41_{-9.94}^{+9.36}$&$243.38_{-13.41}^{+14.36}$&\cite{feng23}\\
                    29.2&HG3&MIRI+NEAR&\texttt{orvara}&$6.31_{-0.56}^{+0.60}$&---&$0.40_{-0.18}^{+0.15}$&$103.7\pm
                                                                                       2.3$&---&\cite{matthews24}\\
      29.2&HG23&---&F23&${5.8}_{-1.0}^{+1.0}$&${162}_{-40}^{+59}$&${0.470}_{-0.10}^{+0.081}$&${99}_{-10}^{+11}$&${234.5}_{-7.3}^{+6.1}$&Solution A\\
      29.2&HG23&MIRI+NEAR&F23&${7.29}_{-0.61}^{+0.60}$&${180}_{-31}^{+32}$& ${0.399}_{-0.076}^{+0.059}$&${105.4}_{-2.4}^{+2.5}$&${228.3}_{-1.6}^{+1.7}$&Solution B\\
      \hline
\multicolumn{10}{l}{$^a$ $T_{rv}$ is the time span of all RV data sets. }\\
\multicolumn{10}{l}{$^b$ The algorithm is largely based on \texttt{orvara} \citep{philipot23a}. }\\
    \end{tabular}
    \label{tab:epsIndA}
  \end{table}

As shown in Table \ref{tab:epsIndA}, the RV time span (\(\Delta T_{rv}\)) is crucial for enhancing the accuracy of predictions derived from combined analyses of RV and astrometry. Solution A provides a predicted location for $\epsilon$ Ind A b consistent with observations, whereas \cite{philipot23a} and F23 (with a 24.8-year RV baseline) yield shorter-period solutions due to the limited RV time span (without HARPSpost2 data), which does not fully capture the RV variation's turn-over (see panel (a) of Fig. \ref{fig:HD209100_rvhg}). When the RV baseline is short, the posterior sampler may also favor a shorter-period solution, interpreting red noise in the RV data as orbital curvature.

The comparison of the companion's position, as predicted by F23 with a 24.8-year RV baseline, Solution A, and Solution B, against the true position observed by JWST/MIRI, is illustrated in Fig.~\ref{fig:HD209100_rel}. For simplicity, the solution provided by \cite{philipot23a} is not included in Fig.~\ref{fig:HD209100_rel}, though their predicted position is similar to that of F23.
As demonstrated in Fig.~\ref{fig:HD209100_rel} and summarized in Table~\ref{tab:epsIndA}, the discrepancy between the predicted and observed position of $\epsilon$~Ind~A~b is independent of the specific algorithm used (F19, F23, or \texttt{orvara}). This is particularly relevant for extremely long-period companions, where the RV data's acceleration derivative must be substantial to resolve degeneracies among \(P\), \(e\), and \(m\). As shown in Table \ref{tab:epsIndA}, both \texttt{orvara} and F23 give consistent orbital solutions if using RV data with a 30-year baseline. 

Compared to the solution provided by \citet{matthews24} and solution A, solution B incorporates all available RV data along with the HG23 data, resulting in an orbital solution with higher precision (see Table \ref{tab:epsIndA}). Therefore, we recommend Solution B for future studies of this planet. The corresponding parameter table and corner plot are presented in Table \ref{tab:pars2} and Fig. \ref{fig:corner_HD209100} in the appendix.

\section{Conclusion}\label{sec:conclusion}
In this work, we demonstrate that the F19 method is similar to \texttt{orvara}, with the difference being that the former models astrometric jitter a posteriori, while the latter uses calibrated HGCA data. Compared to F19 and \texttt{orvara}, the F23 method offers the advantage of incorporating the parallax contribution to GOST-generated abscissae and utilizing multiple Gaia data releases. This approach better constrains the orbits of short-period companions and resolves the degeneracy in $I$ and $\Omega$.

We revisit the orbital solution of HD 28185 using the method developed in \cite{feng23} and applied in \cite{xiao24}. By incorporating the astrometric contribution from the inner planet, HD 28185 b, with its year-long orbit, we derive the inclination and absolute mass of the inner companion. This approach provides a better fit to the astrometric data from Hipparcos, Gaia DR2, and DR3 compared to V24's model that considers only the astrometric signal from the outer companion. This demonstrates the importance of simultaneously modeling both short-period and long-period companions in multi-companion systems, as well as the necessity of accounting for parallax in Gaia and Hipparcos data analysis when constraining year-long orbits. The discrepancy between F22 and V24 in the solution for HD 28185 c is likely due to the use of different RV data sets and insufficient posterior sampling.

We address V24's concerns regarding discrepancies between the
solutions provided by F22 and other studies for HD 28185, HD 38529, 14
Her, HD 62364, GJ 229 B, HD 211847, GJ 680, and HD 111031. These
discrepancies arise mainly for four reasons. First, the use different conventions result in different values of $\Omega$ and $I$ (e.g., HD 38529). In particular, the first and third conventions defined in \cite{feng19c} differs by 180\,deg. Second, insufficient sampling of the double-peaked
inclination posterior distribution leads to the discrepancy in the inclination given by various studies for 14 Her. This issue affects both F22 and many studies using
\texttt{orvara}. Third, the multiple companions would complicate the analyses of RV and astrometric data. The solutions with and without considering potential companions may lead to discrepancies (e.g., GJ 229 B and HD 62364). Fourth, 
discrepancies in orbital periods are
primarily due to differences in the time span of RV data sets
(e.g., HD 28185 and HD 211847). The estimation of periods for wide-orbit companions is highly sensitive to the RV baseline, and this type of discrepancy is data-dependent, regardless of the method used. Underestimations of companion mass for extremely long-period companions (e.g., GJ 229 B, HD 211847 B, GJ 680 B, and HD 111031 B) occur when using only RV and astrometry without incorporating relative astrometry from imaging. This leads to a degeneracy among orbital period, companion mass, and eccentricity, and can result in misinterpreting RV jitter as reflex motion. We reanalyze these systems with imaged stellar companions by modeling their photocentric motion and achieve consistent solutions using the F23 and \texttt{orvara} methods.

Using $\epsilon$ Ind A b as an example, we further emphasize the critical role of the RV baseline in precisely constraining companion mass and orbit. We compare solutions obtained with older and newer RV data, with and without using relative astrometry from JWST/MIRI and the imaging of $\epsilon$ Ind A b conducted by the New Earths in the $\alpha$ Cen Region experiment (NEAR). We conclude that discrepancies between various solutions are primarily due to partial RV coverage of the orbital phase, rather than differences in the methods used. A sufficient RV baseline is crucial for accurately estimating the mass and orbital parameters of companions with decades-long orbits.

Based on our investigation of the discrepancies and issues raised by V24, the following lessons are pertinent for future detections of long-period planets. First, caution is needed when reporting mass and orbital parameters for companions with decades-long orbits, especially when the RV coverage spans less than one-quarter of the full orbital period. However, if relative astrometry is available, the RV baseline can be shorter, provided the acceleration in the RV data is still significant. Second, adequate posterior sampling with multiple samplers using different initial parameters is essential to fully explore multi-modal posteriors and achieve consistent solutions (e.g., \citealt{jin24}). Third, the inclination degeneracy may be resolved by using multiple Gaia data releases, which is crucial for precisely determining the mutual inclination between multiple giant companions in a system. Finally, the conventions for orbital parameters should be clearly defined, as this is essential for future comparative and statistical studies.

Despite existing discrepancies in the solutions for long-period companions, the combined use of RV and astrometry has successfully guided the direct imaging of several giant planets, including AF Lep b \citep{franson23}, HIP 99770 \citep{currie23}, and eps Ind A b \citep{matthews24}. Therefore, we are optimistic about the future synergy between RV, astrometry, and direct imaging, particularly with the expected release of Gaia epoch data in DR4.

\section*{Acknowledgments}
We sincerely thank the anonymous referee for their insightful comments and suggestions, which have greatly improved the quality and clarity of this manuscript. We sincerely thank Markus Janson and Johanna Teske for their valuable suggestions. This work is supported by the National Key R\&D Program of China, No. 2024YFA1611801 and No. 2024YFC2207800, by the National Natural Science Foundation of China (NSFC) under Grant No. 12473066, by the Shanghai Jiao Tong University 2030 Initiative, and by the China-Chile Joint Research Fund (CCJRF No. 2205). CCJRF is provided by Chinese Academy of Sciences South America Center for Astronomy (CASSACA) and established by National Astronomical Observatories, Chinese Academy of Sciences (NAOC) and Chilean Astronomy Society (SOCHIAS) to support China-Chile collaborations in astronomy. This work is partly based on observations collected at the European Organisation for Astronomical Research in the Southern Hemisphere under ESO programs: 108.22KV, 60.A-9036, 072.C-0488, 079.C-0681, 087.D-0511, 091.C-0844, 094.C-0894, 192.C-0852, 196.C-1006, 105.20MP.001, 108.2271.002 and 108.2271.003.

\section*{Data Availability}
The RV and imaging data are available in the literature, whereas the Gaia and Hipparcos data are publicly accessible.

\bibliographystyle{mnras}
\bibliography{nm}

\appendix
\section{Parameters of companions}

\begin{landscape}
\begin{table}
\renewcommand{\arraystretch}{1.5}
\centering
\caption{Parameters for GJ\,229\,B, HD\,62364\,B, HD\,38529\,c, and 14\,Her system. Stellar mass:  HD\,62364 ($1.19\pm0.16\,M_{\odot}$,\citealt{feng22}), HD\,38529 ($1.36\pm0.02\,M_{\odot}$, \citealt{xuan20}), 14\,Her ($0.98\pm0.04\,M_{\odot}$, \citealt{bardalez21}).}\label{tab:pars1}
\begin{tabular}{lcccccc}
\hline \hline 
Parameter$^{\rm a}$ & GJ\,229\,B$^{\rm c}$ & HD\,62364\,B& \multicolumn{2}{c}{HD\,38529\,c}&14\,Her\,b&14\,Her\,c\\
 &F23& F23 & \texttt{orvara} (DR2) &\texttt{orvara} (EDR3)&\multicolumn{2}{c}{\texttt{orvara}}\\
\hline
\textbf{Fitted Parameter}&\\
Orbital period $P$ (day)&${98863}_{-2245}^{+2348}$&${5163}_{-20}^{+21}$&${2127.8}_{-3.3}^{+3.1}$&${2127.8}_{-3.2}^{+3.3}$&${1763.63}_{-0.84}^{+0.80}$&${51748}_{-18353}^{+33266}$\\
RV semi-amplitude $K$ (m s$^{-1}$) &${375}_{-12}^{+11}$&${170.3}_{-2.2}^{+2.2}$&${169.8}_{-1.1}^{+1.1}$&${169.8}_{-1.1}^{+1.1}$&${90.32}_{-0.42}^{+0.41}$&${44.2}_{-2.5}^{+2.7}$\\
Eccentricity $e$&${0.8121}_{-0.0054}^{+0.0044}$&${0.6092}_{-0.0044}^{+0.0045}$&${0.3505}_{-0.0053}^{+0.0055}$&${0.3507}_{-0.0051}^{+0.0057}$&${0.3698}_{-0.0034}^{+0.0032}$&${0.644}_{-0.11}^{+0.092}$\\
Argument of periapsis $\omega$ (deg)&${325.0}_{-1.0}^{+1.0}$&${-0.26}_{-0.77}^{+0.76}$&${22.3}_{-1.4}^{+1.3}$&${22.1}_{-1.3}^{+1.3}$&${22.79}_{-0.48}^{+0.48}$&${14.6}_{-5.5}^{+5.6}$\\
Mean anomaly$^{\rm b}$ $M_0$ (deg)&${322.92}_{-0.64}^{+0.64}$&${44.8}_{-1.4}^{+1.4}$&${134.1}_{-1.1}^{+1.1}$&${134.2}_{-1.1}^{+1.0}$& ${59.91}_{-0.39}^{+0.38}$&${20.6}_{-8.3}^{+12}$\\
Inclination $I$ (deg)&${31.1}_{-1.5}^{+1.7}$&${133.1}_{-1.3}^{+1.3}$&${129.3}_{-14}^{+9.4}$&${104.2}_{-11}^{+8.9}$&${33.1}_{-3.8}^{+6.5}$ (${144.0}_{-8.8}^{+7.3}$)& ${100}_{-33}^{+32}$ (${135}_{-26}^{+10}$)\\
Longitude of ascending node $\Omega$ (deg)&${170.42}_{-0.47}^{+0.46}$&${98.4}_{-3.0}^{+3.0}$&${218}_{-15}^{+16}$&${238}_{-17}^{+37}$&${236}_{-14}^{+14}$ (${35}_{-12}^{+13}$) & ${311}_{-60}^{+32}$ (${26}_{-17}^{+301}$)\\\hline
\textbf{Derived Parameter}&\\
Orbital period $P$ (yr)&${270.7}_{-6.1}^{+6.4}$&${14.135}_{-0.056}^{+0.057}$&${5.8256}_{-0.0090}^{+0.0085}$&${5.8256}_{-0.0087}^{+0.0091}$&${4.8285}_{-0.0023}^{+0.0022}$&${142}_{-50}^{+91}$\\
Semi-major axis $a$ (au)&${36.10}_{-0.47}^{+0.47}$&${6.236}_{-0.071}^{+0.070}$&${3.607}_{-0.039}^{+0.039}$&${3.604}_{-0.042}^{+0.043}$&${2.843}_{-0.040}^{+0.040}$&${27.0}_{-6.8}^{+11}$\\
Companion mass $m$ ($M_{\rm Jup}$)&${71.73}_{-0.56}^{+0.55}$&${17.90}_{-0.53}^{+0.54}$&${16.2}_{-2.3}^{+2.9}$&${12.93}_{-0.49}^{+0.70}$&${8.9}_{-1.5}^{+1.1}$&${7.3}_{-1.1}^{+2.3}$\\
Periapsis epoch $T_{\rm p}-2400000$ (JD)&${-38253}_{-2278}^{+2177}$&${52405}_{-23}^{+23}$&${54405.2}_{-6.3}^{+6.4}$&${54404.4}_{-5.9}^{+6.3}$&${54904.0}_{-1.9}^{+1.9}$&${52224}_{-158}^{+159}$\\
\hline
\textbf{Barycentric Offset}&\\
$\alpha*$ offset $\Delta \alpha*$ (mas)&${-52.18}_{-0.65}^{+0.64}$&${0.117}_{-0.031}^{+0.031}$&---&---&---&---\\
$\delta$ offset $\Delta \delta$ (mas)&${594.9}_{-7.3}^{+7.3}$&${-0.282}_{-0.028}^{+0.029}$&---&---&---&---\\
$\mu_{\alpha*}$ offset $\Delta \mu_{\alpha*}$ (mas\,yr$^{-1}$)&${9.89}_{-0.12}^{+0.12}$&${0.439}_{-0.016}^{+0.017}$&---&---&---&---\\
$\mu_\delta$ offset $\Delta \mu_\delta$ (mas\,yr$^{-1}$)&${-13.47}_{-0.14}^{+0.14}$&${0.673}_{-0.016}^{+0.016}$&---&---&---&---\\
$\varpi$ offset $\Delta \varpi$ (mas)&${-0.036}_{-0.018}^{+0.018}$&${0.021}_{-0.022}^{+0.022}$&---&---&---&---\\
\hline
\multicolumn{7}{l}{$^{\rm a}$ The fitted parameters used by \texttt{orvara} differ from ours, so we convert their values to the same form using MCMC posteriors. }\\
\multicolumn{7}{l}{$^{\rm b}$ The reference epoch for our model is set to the minimum Julian day of RVs, while for \texttt{orvara}, it is 2010.0 yr or JD = 2455197.50. }\\
\multicolumn{7}{l}{$^{\rm c}$ We adopt an uniform prior for the stellar mass of GJ\,229. The best-fit value is $0.576\pm0.007\,M_{\odot}$, comparable with the value of $0.579\pm0.007\,M_{\odot}$ given by \citet{brandt21d}.}\\
\end{tabular}
\end{table}
\end{landscape}

\begin{landscape}
\begin{table}
\renewcommand{\arraystretch}{1.5}
\centering
\caption{Parameters for eps Ind A b, HD 211847 B, HD111031 B, and GJ 680 B. Stellar mass: eps Ind A ($0.76\pm0.04\,M_{\odot}$, \citealt{matthews24}), HD\,211847 ($0.94\pm0.04\,M_{\odot}$,\citealt{sahlmann11}, HD\,111031 ($1.17\pm0.06\,M_{\odot}$, \citealt{kervella22}), GJ\,680 ($0.47\pm0.01\,M_{\odot}$, \citealt{kervella22}).}\label{tab:pars2}
\begin{tabular}{lcccccccc}
\hline \hline 
Parameter$^{\rm a}$ & \multicolumn{2}{c}{eps Ind A b} & \multicolumn{2}{c}{HD\,211847\,B}& \multicolumn{2}{c}{HD\,111031\,B}&\multicolumn{2}{c}{GJ\,680\,B}\\
 & Solution A (no DI) & Solution B (with DI)&F23&\texttt{orvara}&F23&\texttt{orvara}&F23&\texttt{orvara}\\
\hline
\textbf{Fitted Parameter}&\\
Orbital period $P$ (day)&${59001}_{-14428}^{+21636}$&${65692}_{-11189}^{+11580}$&${6201}_{-17}^{+16}$&${6205}_{-18}^{+19}$&${49618}_{-3276}^{+4339}$&${51160}_{-3745}^{+4430}$&${95842}_{-1339}^{+1139}$&${143676}_{-43902}^{+113121}$\\
RV semi-amplitude $K$ (m s$^{-1}$) &${39.6}_{-4.8}^{+4.8}$&${45.6}_{-3.2}^{+3.4}$&${275}_{-16}^{+19}$&${254.3}_{-7.6}^{+7.2}$&${543}_{-24}^{+24}$&${550}_{-33}^{+35}$&${986}_{-13}^{+13}$&${896}_{-116}^{+125}$\\
Eccentricity $e$&${0.470}_{-0.10}^{+0.081}$&${0.399}_{-0.076}^{+0.059}$&${0.5565}_{-0.010}^{+0.0095}$&${0.563}_{-0.011}^{+0.011}$&${0.541}_{-0.023}^{+0.024}$&${0.531}_{-0.037}^{+0.033}$&${0.0479}_{-0.0067}^{+0.0075}$&${0.42}_{-0.23}^{+0.20}$\\
Argument of periapsis $\omega$ (deg)&${28}_{-15}^{+16}$&${17.8}_{-8.8}^{+12}$&${165.5}_{-4.2}^{+4.8}$&${171.1}_{-1.9}^{+3.2}$& ${82.0}_{-3.7}^{+5.1}$&${83.4}_{-5.6}^{+6.4}$&${185.7}_{-4.7}^{+4.2}$&${263}_{-43}^{+37}$\\
Mean anomaly$^{\rm b}$ $M_0$ (deg)&${303}_{-22}^{+17}$&${306}_{-12}^{+10}$&${263.1}_{-1.3}^{+1.2}$&${31.3}_{-3.2}^{+3.3}$&${58.44}_{-1.0}^{+0.89}$&${65.5}_{-5.6}^{+5.6}$&${161.2}_{-4.6}^{+5.1}$&${41}_{-25}^{+57}$\\
Inclination $I$ (deg)&${99}_{-10}^{+11}$&${105.4}_{-2.4}^{+2.5}$&${171.29}_{-0.65}^{+0.57}$&${172.04}_{-0.36}^{+0.32}$&${147.4}_{-1.3}^{+1.3}$&${147.6}_{-1.7}^{+1.7}$&${115.48}_{-0.94}^{+1.0}$&${123.8}_{-8.9}^{+10}$\\
Longitude of ascending node$^{\rm c}$ $\Omega$ (deg)&${234.5}_{-7.3}^{+6.1}$&${228.3}_{-1.6}^{+1.7}$&${-3.5}_{-4.2}^{+4.8}$&${2.5}_{-1.8}^{+3.3}$&${160.0}_{-1.2}^{+1.2}$&${160.2}_{-1.6}^{+1.7}$&${140.61}_{-0.32}^{+0.32}$&${139.1}_{-2.0}^{+1.6}$\\\hline
\textbf{Derived Parameter}&\\
Orbital period $P$ (yr)&${162}_{-40}^{+59}$&${180}_{-31}^{+32}$&${16.978}_{-0.045}^{+0.045}$&${16.990}_{-0.049}^{+0.051}$&${135.8}_{-9.0}^{+12}$&${140}_{-10}^{+12}$&&${393}_{-120}^{+310}$\\
Semi-major axis $a$ (au)&${27.0}_{-4.5}^{+6.6}$&${29.2}_{-3.4}^{+3.3}$&${6.567}_{-0.062}^{+0.063}$&${6.839}_{-0.061}^{+0.061}$&${27.53}_{-0.88}^{+1.2}$&${29.2}_{-1.2}^{+1.4}$&${32.05}_{-0.31}^{+0.29}$&${46}_{-10}^{+22}$\\
Companion mass $m$ ($M_{\rm Jup}$)&${5.8}_{-1.0}^{+1.0}$&${7.29}_{-0.61}^{+0.60}$&${147.8}_{-3.1}^{+3.1}$&${146.7}_{-3.3}^{+3.4}$&${183.5}_{-5.2}^{+5.4}$&${186.9}_{-6.7}^{+7.2}$&${185.5}_{-3.2}^{+2.7}$&${192.3}_{-4.6}^{+4.4}$\\
Periapsis epoch $T_{\rm p}-2400000$ (JD)&${-647}_{-21953}^{+14940}$&${-6945}_{-11673}^{+11272}$&${48016}_{-29}^{+30}$&${54190}_{-16}^{+18}$&${46133}_{-251}^{+241}$&${45910}_{-384}^{+332}$&${10235}_{-1238}^{+1370}$&${38512}_{-7891}^{+4231}$\\
\hline
\textbf{Barycentric Offset}&\\
$\alpha*$ offset $\Delta \alpha*$ (mas)&${-23.5}_{-7.6}^{+6.6}$&${-33.1}_{-3.5}^{+3.5}$&${4.53}_{-0.24}^{+0.25}$&---&${-114.7}_{-4.6}^{+4.1}$&---&${677.9}_{-12}^{+9.1}$&---\\
$\delta$ offset $\Delta \delta$ (mas)&${-17.2}_{-7.5}^{+5.3}$&${-29.0}_{-3.1}^{+2.7}$&${26.79}_{-0.58}^{+0.60}$&---&${77.2}_{-3.4}^{+3.6}$&---&${-787}_{-12}^{+12}$&---\\
$\mu_{\alpha*}$ offset $\Delta \mu_{\alpha*}$ (mas\,yr$^{-1}$)&${0.77}_{-0.39}^{+0.36}$&${0.76}_{-0.15}^{+0.16}$&${-3.539}_{-0.044}^{+0.043}$&---&${-3.36}_{-0.21}^{+0.17}$&---&${6.89}_{-0.30}^{+0.29}$&---\\
$\mu_\delta$ offset $\Delta \mu_\delta$ (mas\,yr$^{-1}$)&${0.21}_{-0.55}^{+0.41}$&${-0.457}_{-0.087}^{+0.085}$&${0.421}_{-0.029}^{+0.029}$&---&${-2.12}_{-0.16}^{+0.17}$&---&${7.14}_{-0.24}^{+0.27}$&---\\
$\varpi$ offset $\Delta \varpi$ (mas)&${-0.11}_{-0.11}^{+0.11}$&${-0.11}_{-0.11}^{+0.11}$&${0.014}_{-0.034}^{+0.034}$&---&${0.009}_{-0.025}^{+0.027}$&---&${0.015}_{-0.026}^{+0.026}$&---\\
\hline
\multicolumn{9}{l}{$^{\rm a}$ The fitted parameters used by \texttt{orvara} differ from ours, so we convert their values to the same form using MCMC posteriors. }\\
\multicolumn{9}{l}{$^{\rm b}$ The reference epoch for our model is set to the minimum Julian day of RVs, while for \texttt{orvara}, it is 2010.0 yr or JD = 2455197.50. }\\
\multicolumn{9}{l}{$^{\rm c}$ We shift a $\pi$ from $\Omega$ of \texttt{orvara} to comfort to our adopted coordinate system convention (see Convention III of Appendix A of \citealt{feng19c}). }\\
\end{tabular}
\end{table}
\end{landscape}

\section{Posterior distributions}
\begin{figure*}
    \centering
	\includegraphics[width=\textwidth]{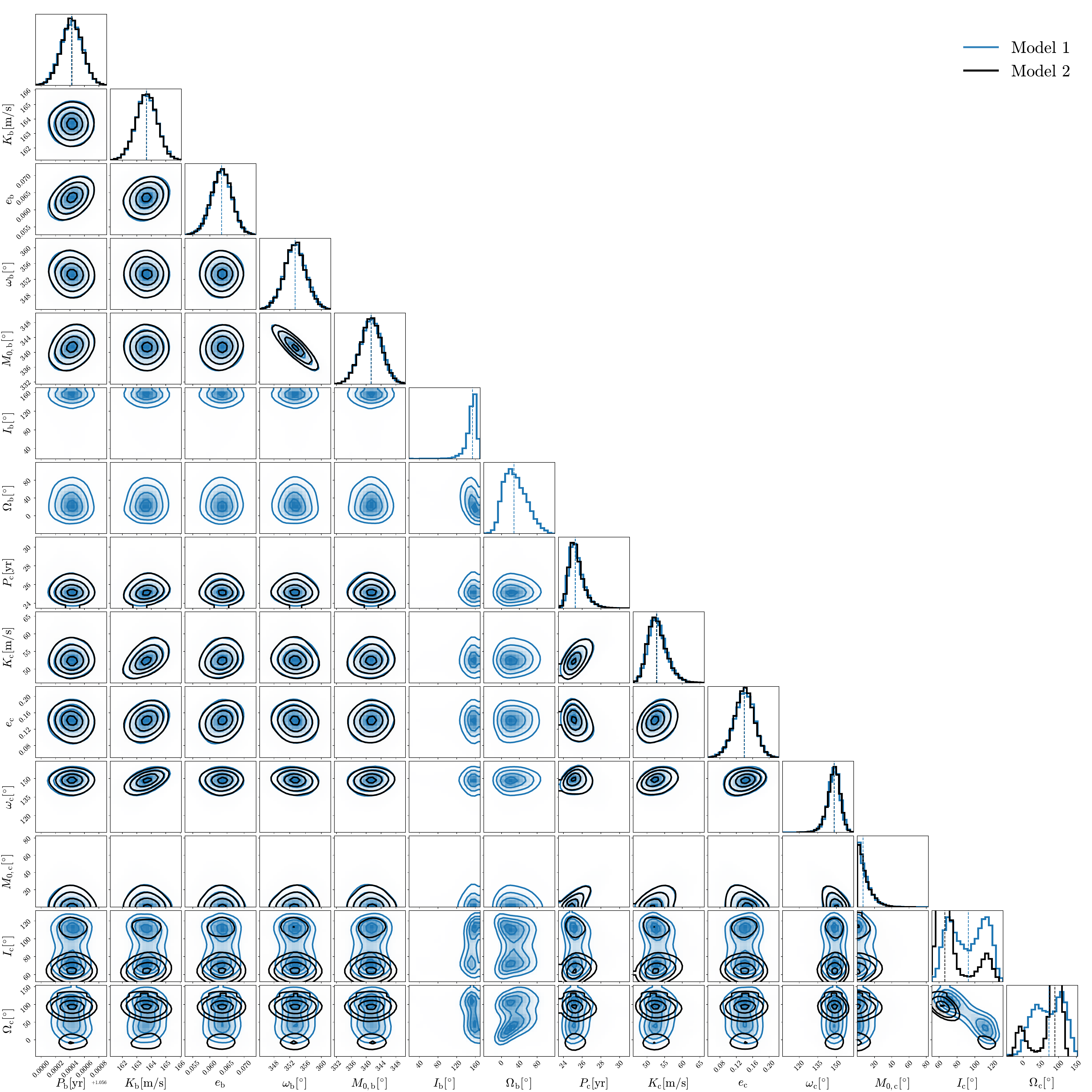}
    \caption{Posterior distributions of the selected orbital parameters for HD\,28185 system. The median is denoted by a vertical dashed line.}
    \label{fig:corner_HD28185}
\end{figure*}

\begin{figure*}
    \centering
	\includegraphics[width=\textwidth]{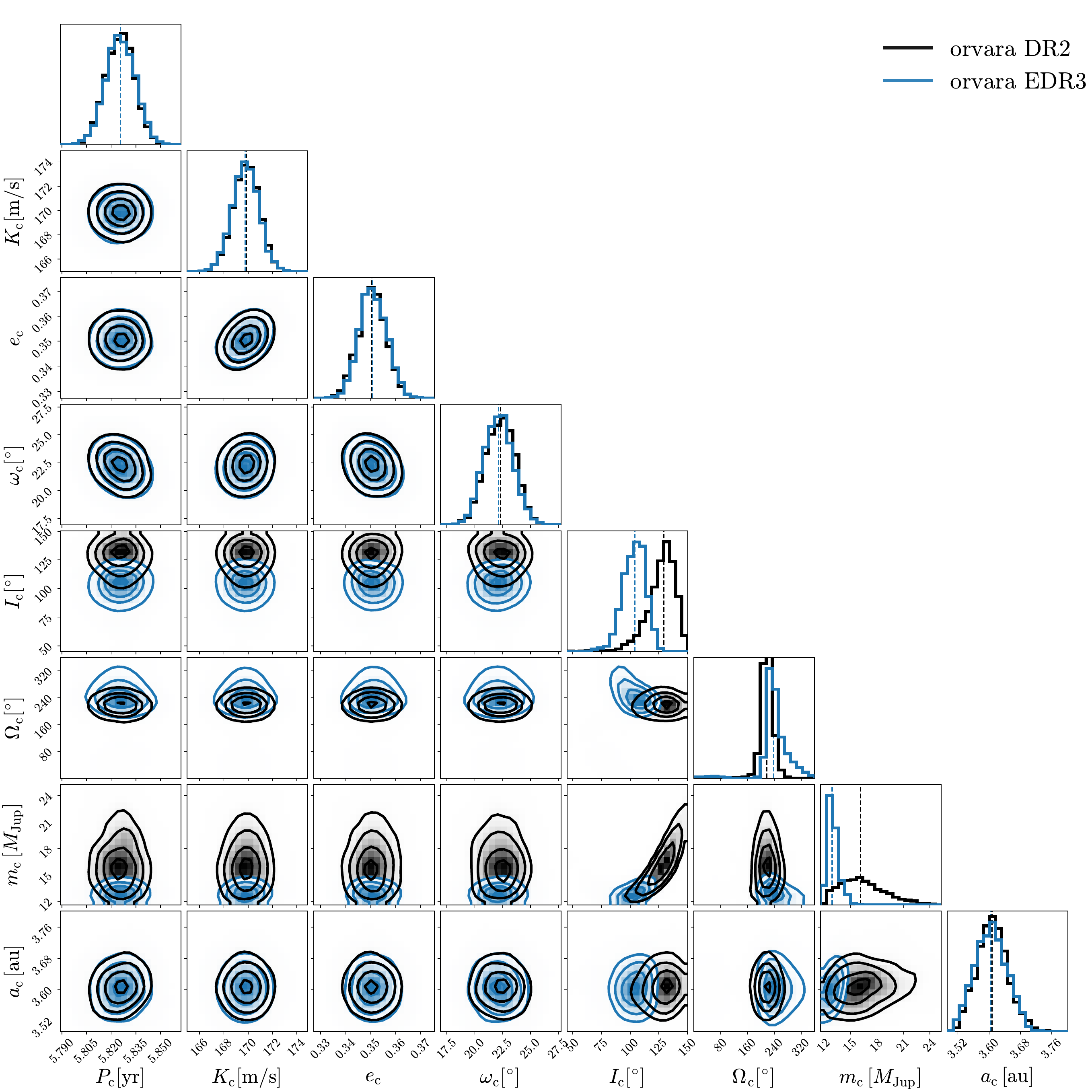}
    \caption{Posterior distributions of the selected orbital parameters for HD\,38529 system. }
    \label{fig:corner_HD38529}
\end{figure*}

\begin{figure*}
    \centering
	\includegraphics[width=\textwidth]{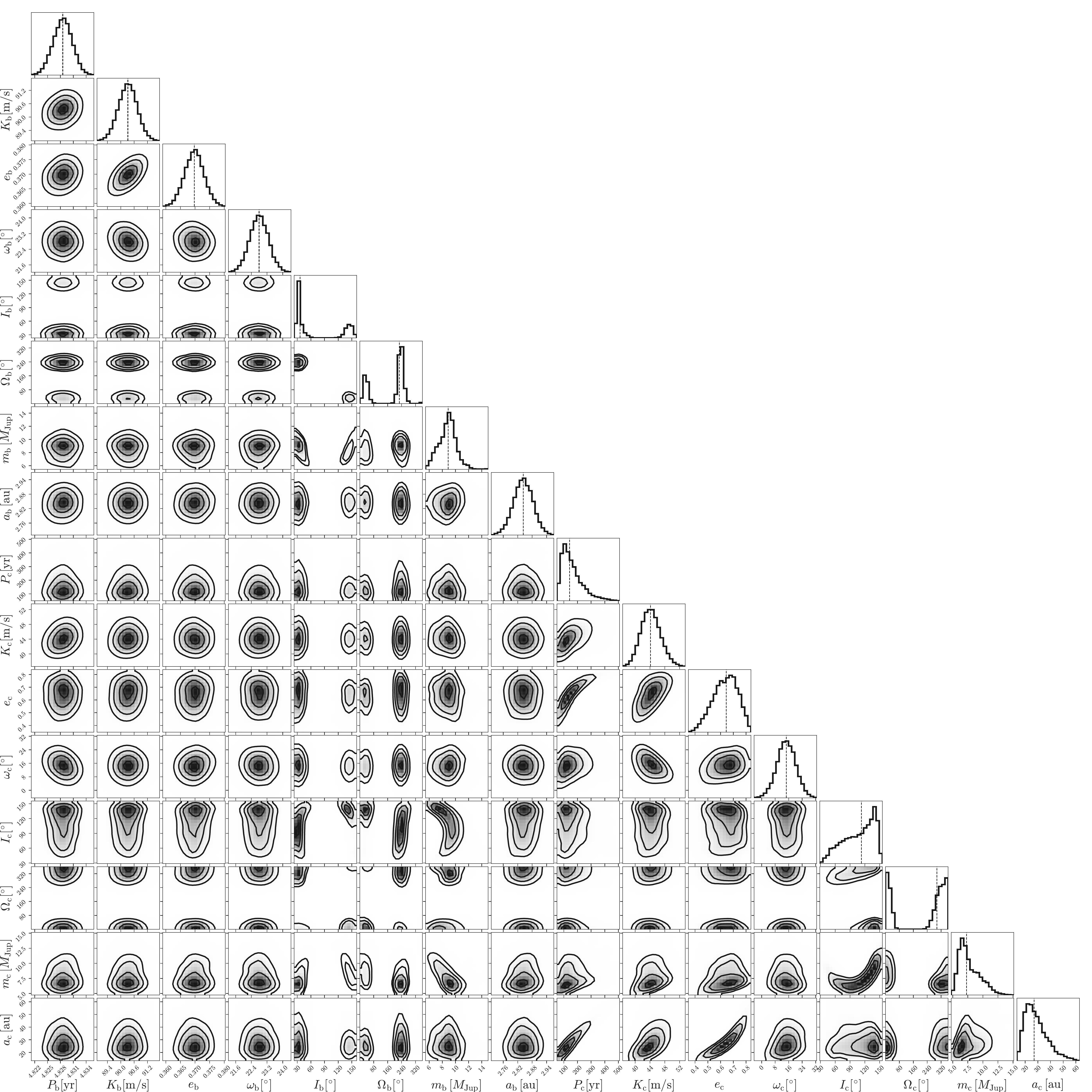}
    \caption{Posterior distributions of the selected orbital parameters for 14\,Her system. }
    \label{fig:corner_14Her}
\end{figure*}

\begin{figure*}
    \centering
	\includegraphics[width=\textwidth]{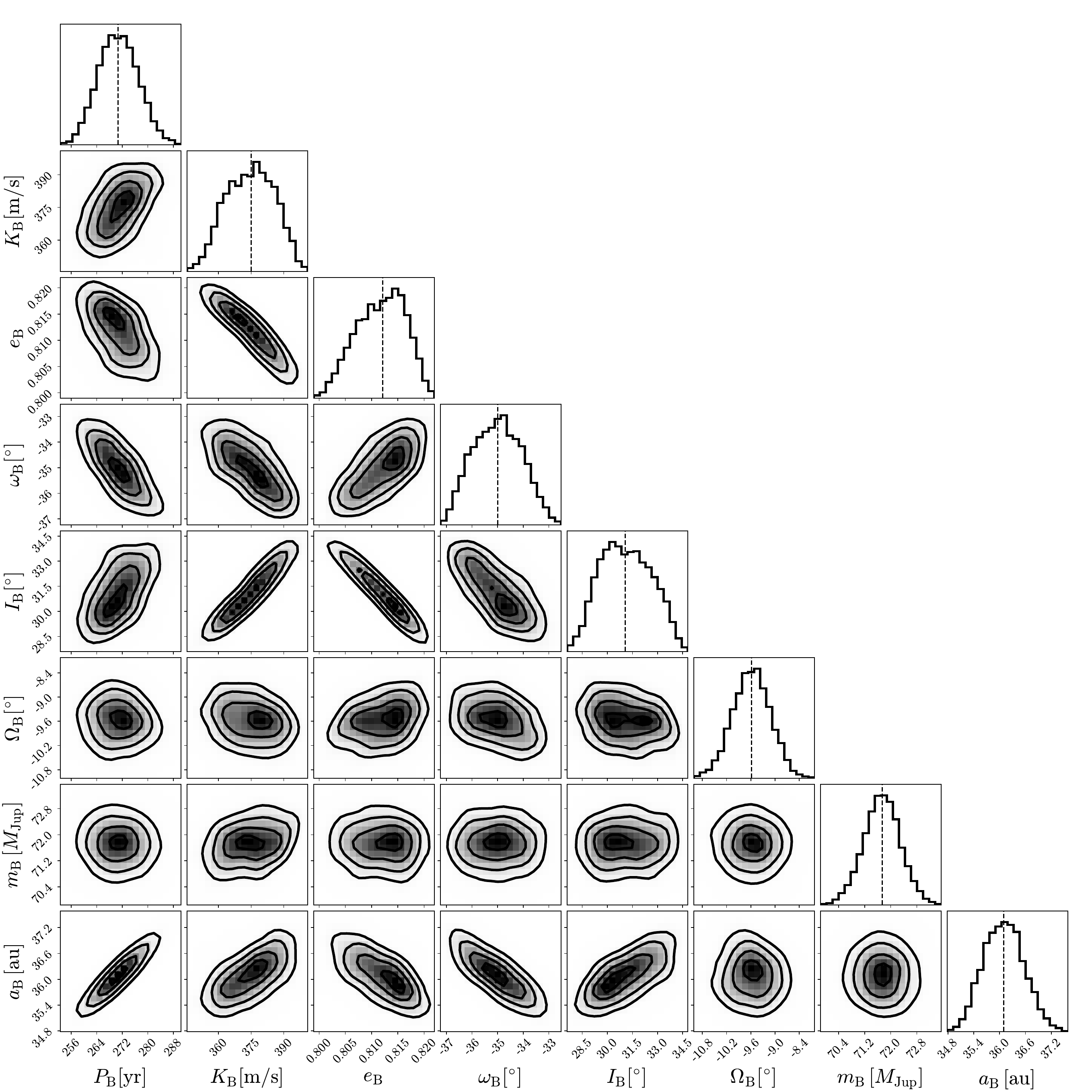}
    \caption{Posterior distributions of the selected orbital parameters for GJ\,229 system. }
    \label{fig:corner_GJ229}
\end{figure*}
\begin{figure*}
    \centering
	\includegraphics[width=\textwidth]{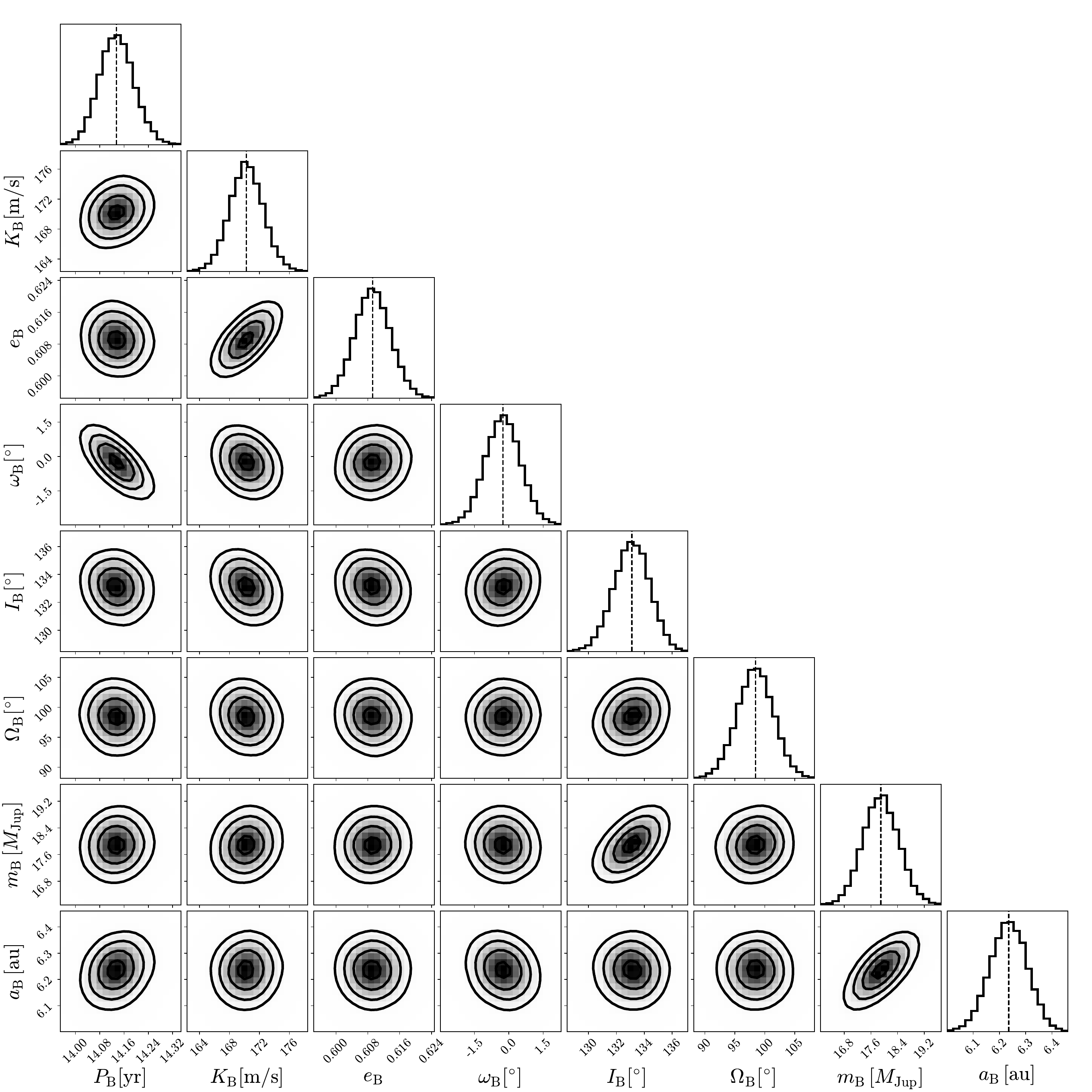}
    \caption{Posterior distributions of the selected orbital parameters for HD\,62364 system. }
    \label{fig:corner_HD62364}
\end{figure*}

\begin{figure*}
    \centering
	\includegraphics[width=\textwidth]{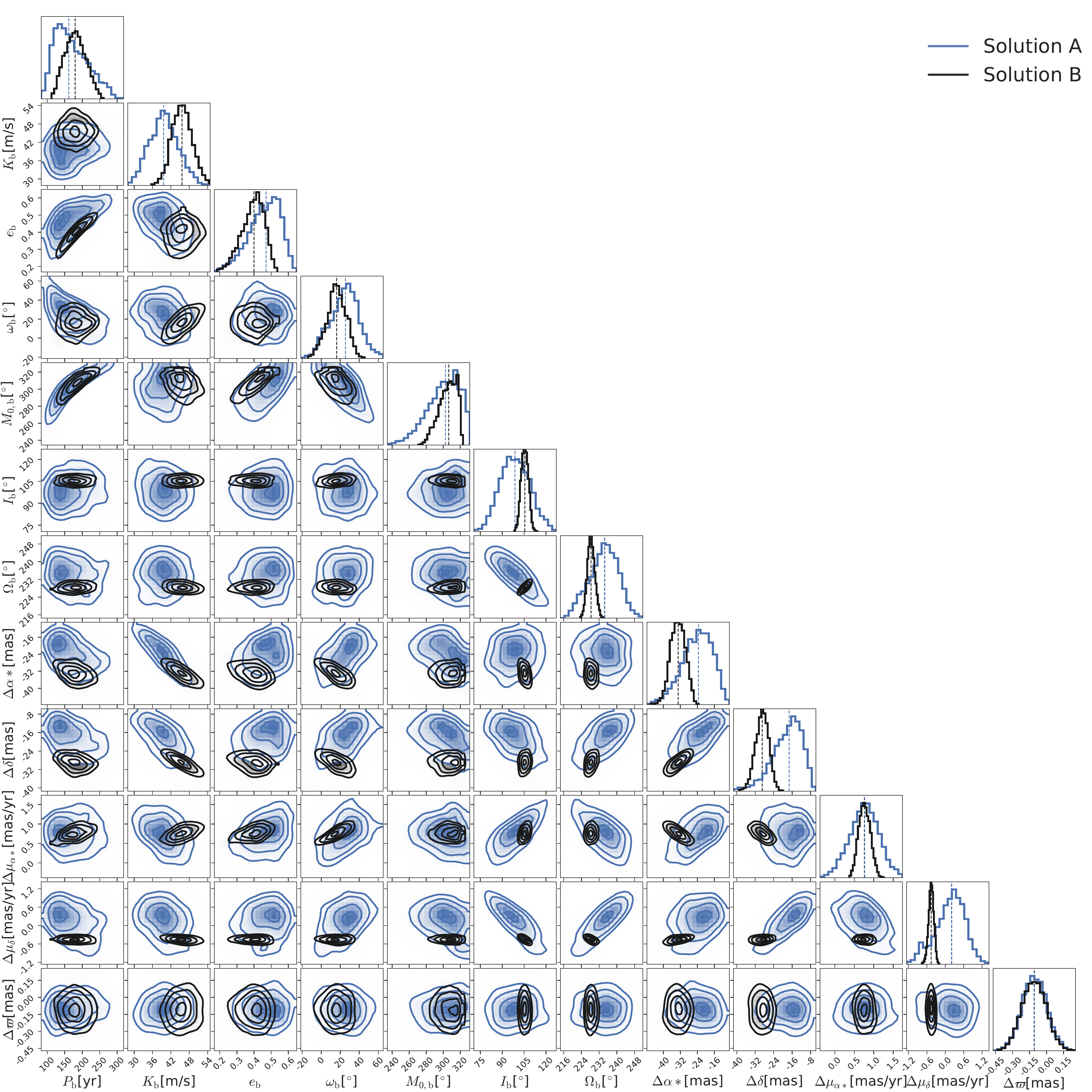}
    \caption{Posterior distributions of the selected orbital parameters for eps Ind A system. }
    \label{fig:corner_HD209100}
\end{figure*}

\begin{figure*}
    \centering
	\includegraphics[width=\textwidth]{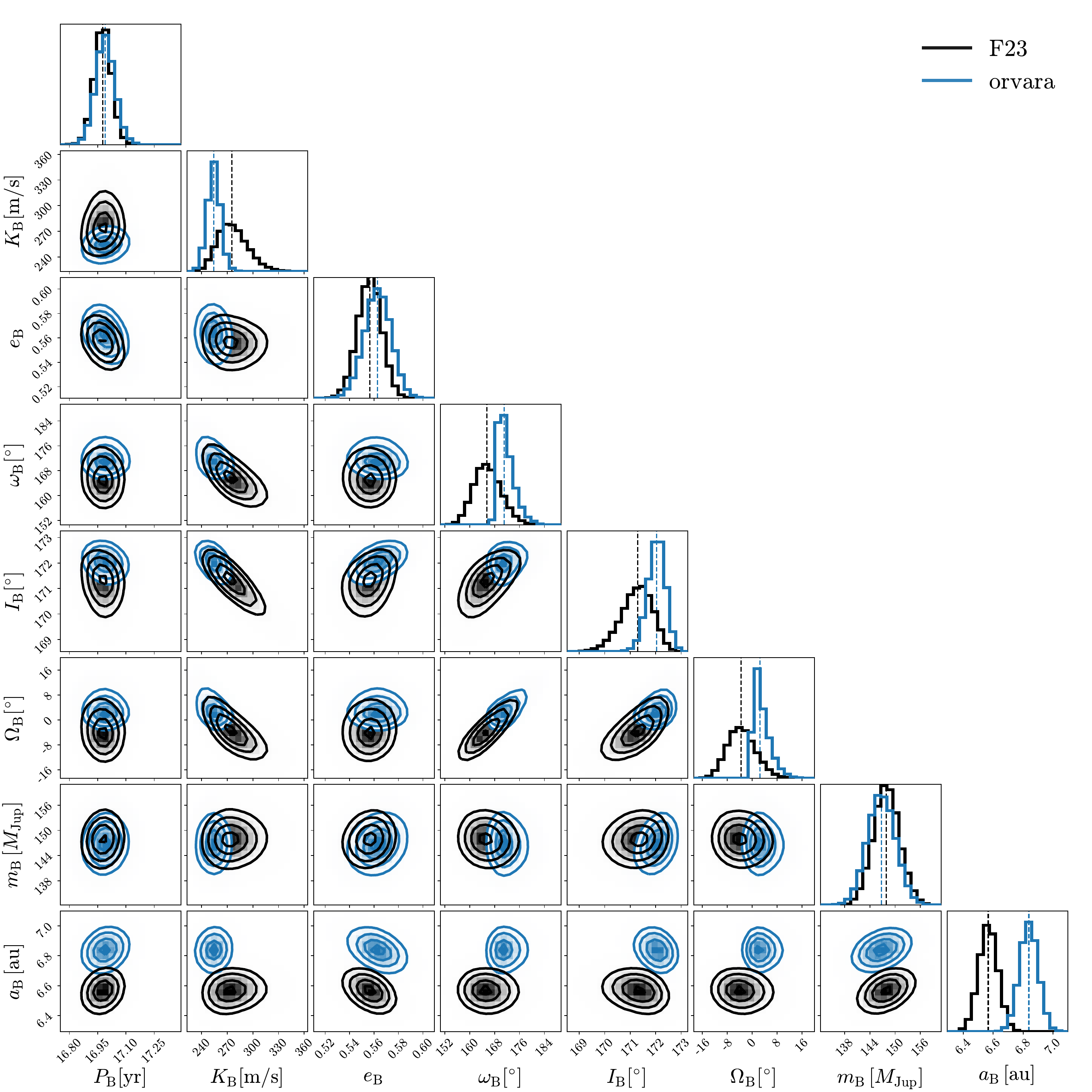}
    \caption{Posterior distributions of the selected orbital parameters for HD\,211847 system. }
    \label{fig:corner_HD211847}
\end{figure*}

\begin{figure*}
    \centering
	\includegraphics[width=\textwidth]{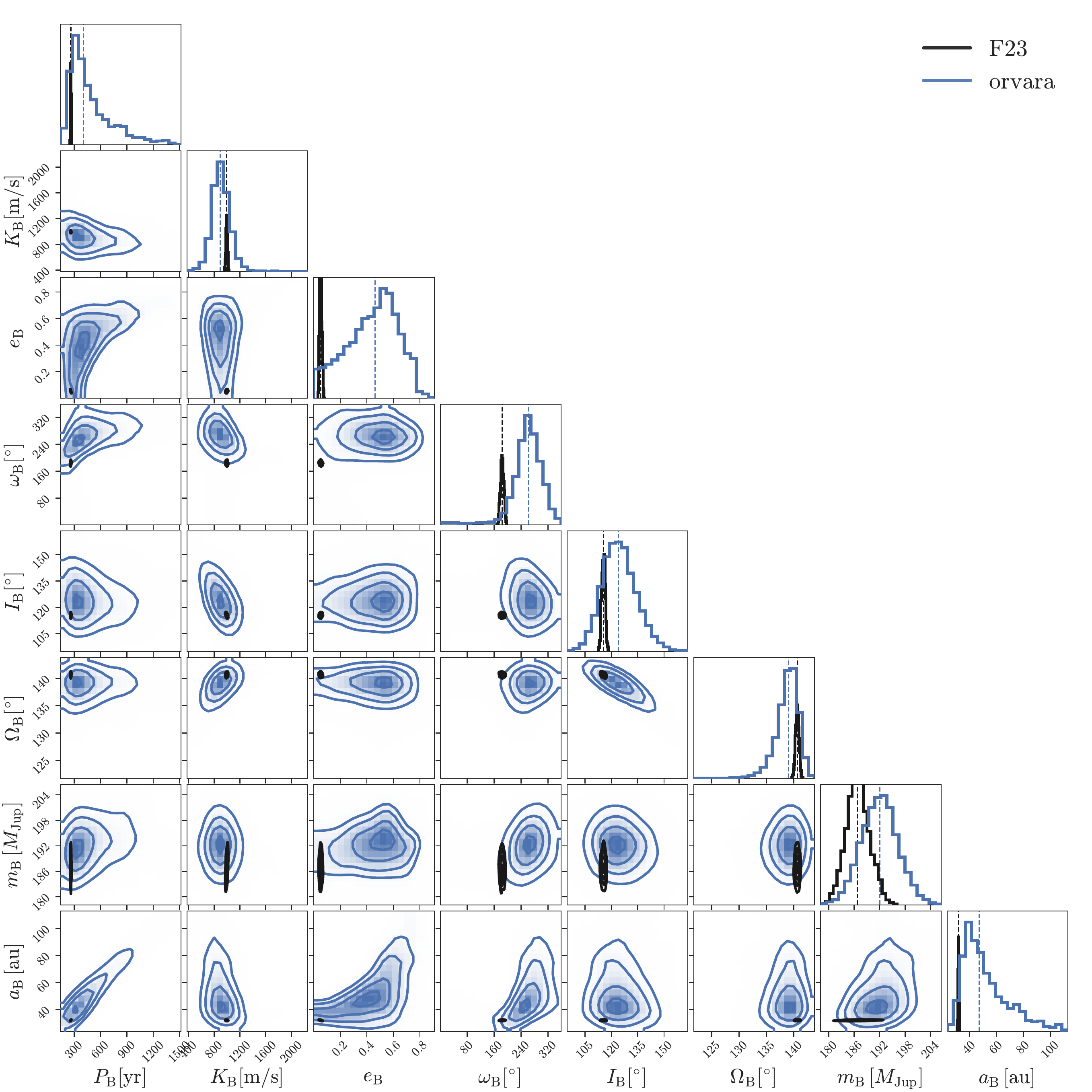}
    \caption{Posterior distributions of the selected orbital parameters for GJ\,680 system. }
    \label{fig:corner_GJ680}
\end{figure*}

\begin{figure*}
    \centering
	\includegraphics[width=\textwidth]{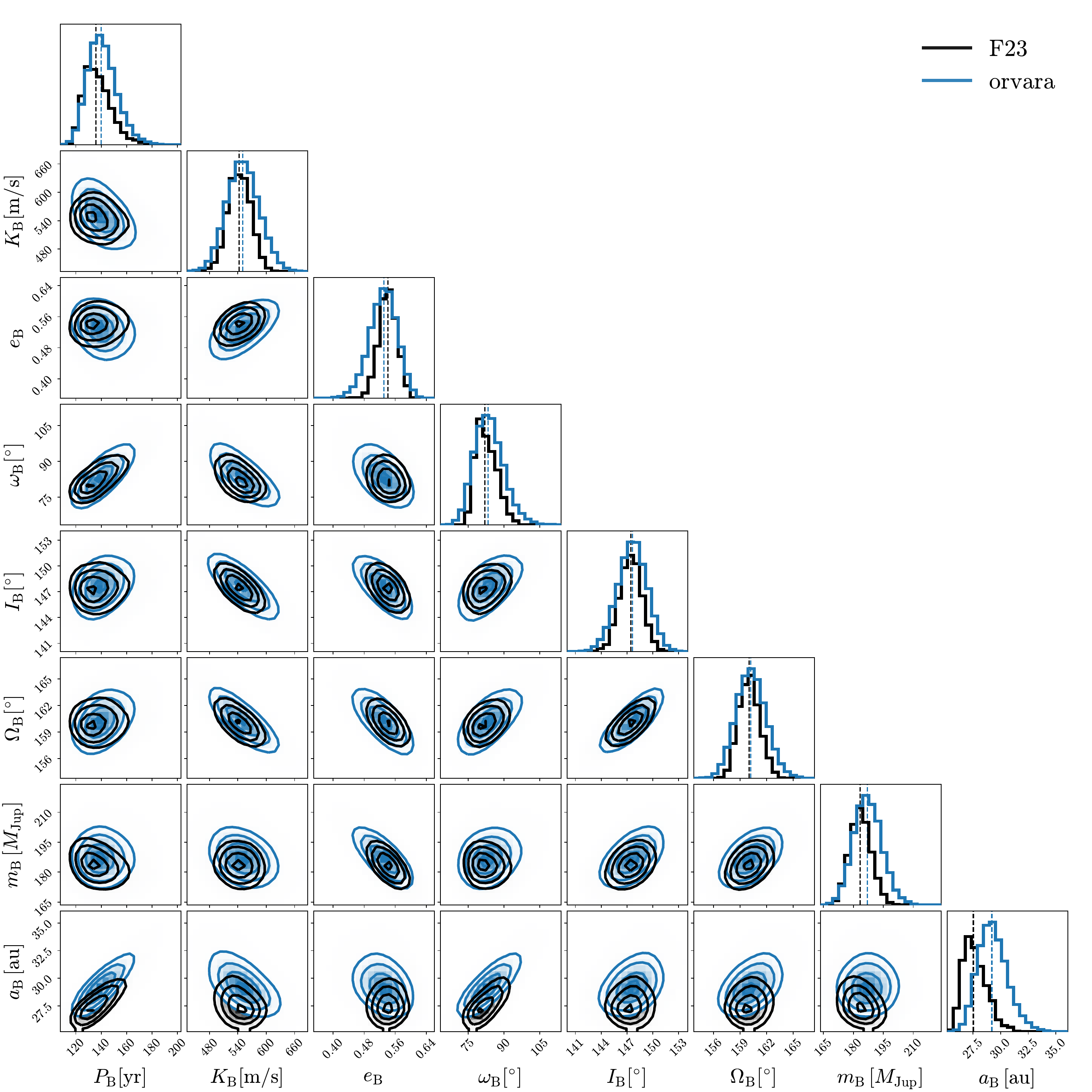}
    \caption{Posterior distributions of the selected orbital parameters for HD\,111031 system. }
    \label{fig:corner_HD111031}
\end{figure*}

\end{document}